\UseRawInputEncoding

\documentclass[journal]{IEEEtran}  %%Regular Journal Article

\usepackage{amsmath,amsfonts}
\usepackage{algorithmic}
\usepackage{array}
\usepackage[caption=false,font=footnotesize,labelfont=rm,textfont=rm]{subfig}
\usepackage{textcomp}
\usepackage{stfloats}
\usepackage{url}
\usepackage{verbatim}
\usepackage{graphicx}

\usepackage{color}
\usepackage{amssymb}
\usepackage{booktabs}
\usepackage[table]{xcolor}
\usepackage{graphicx}
\usepackage{multirow}
\usepackage{color} 
\usepackage{cite}
\newcommand{\figref}[1]{Fig. \ref{#1}}
\newcommand{\figrefs}[1]{Figs. \ref{#1}}
\newcommand{\ra}[1]{\renewcommand{\arraystretch}{#1}}
\newcommand{\subfigref}[2]{Fig. \ref{#1} \subref{#2}}

\def\BibTeX{{\rm B\kern-.05em{\sc i\kern-.025em b}\kern-.08em
    T\kern-.1667em\lower.7ex\hbox{E}\kern-.125emX}}
\usepackage{balance}

\begin{document}
	\title{An I2I Inpainting Approach for Efficient Channel Knowledge Map Construction}

\author{
	Zhenzhou Jin,~\IEEEmembership{Graduate Student Member,~IEEE,}
	Li~You,~\IEEEmembership{Senior Member,~IEEE,}
	Jue~Wang,~\IEEEmembership{Member,~IEEE,}
	%Christos~G. Tsinos,~\IEEEmembership{Senior Member,~IEEE,}\\
	%Fan~Liu,~\IEEEmembership{Member,~IEEE,}
	%Wenjin~Wang,~\IEEEmembership{Member,~IEEE,} \\
	Xiang-Gen~Xia,~\IEEEmembership{Fellow,~IEEE,}
	and~Xiqi~Gao,~\IEEEmembership{Fellow,~IEEE}
	%and~Bj\"{o}rn~Ottersten,~\IEEEmembership{Fellow,~IEEE}
\thanks{Part of this work was presented in the IEEE WCNC 2024 \cite{jin}.}	
\thanks{
		%This work was supported by the National Key Research and Development
%		Program of China under Grant 2018YFB1801103, the Key Technologies
%		R\&D Program of Jiangsu (Prospective and Key Technologies for Industry)
%		under Grants BE2022067 and BE2022067-5, the Jiangsu Province Basic
%		Research Project under Grant BK20192002, the Fundamental Research Funds
%		for the Central Universities under Grant 2242021R41148, and the Young Elite
%		Scientist Sponsorship Program by China Institute of Communications. The
%		work of Jue Wang was supported by the National Natural Science Foundation
%		of China under Grant 62171240. \emph{(Corresponding author: Li You.)}
		Zhenzhou Jin, Li You, and Xiqi Gao are with the National Mobile Communications Research Laboratory, Southeast University, Nanjing 210096, China, and also with the Purple Mountain Laboratories, Nanjing 211100, China (e-mail: zzjin@seu.edu.cn; lyou@seu.edu.cn; xqgao@seu.edu.cn).
		
		Jue Wang is with School of Information Science and Technology, Nantong
		University, Nantong 226019, China, and also with Nantong Research Institute for Advanced Communication Technologies, Nantong 226019, China (e-mail: wangjue@ntu.edu.cn).
		
		Xiang-Gen Xia is with the Department of Electrical and Computer Engineering, University of Delaware, Newark, DE 19716 USA (e-mail: xxia@ee.udel.edu).
	}

%\author{ Zhenzhou Jin, Li You %~\IEEEmembership{Student~Member,~IEEE,}
	%~\IEEEmembership{Fellow,~IEEE}% <-this % stops a space

	%\thanks{This work was supported by the National Key Research and Development
		%Program of China under Grant 2018YFB1801103, the Key Technologies
		%R\&D Program of Jiangsu (Prospective and Key Technologies for Industry)
		%under Grants BE2022067 and BE2022067-5, the Jiangsu Province Basic
		%Research Project under Grant BK20192002, the Fundamental Research Funds
		%for the Central Universities under Grant 2242021R41148, and the Young Elite
		%Scientist Sponsorship Program by China Institute of Communications. The
		%work of Jue Wang was supported by the National Natural Science Foundation
		%of China under Grant 62171240. The associate editor coordinating the review of this paper and approving it for publication was Dr. Yingyang Chen. \emph{(Corresponding author: Li You.)}

		%Xiang Xiao, Li You, Wenjin Wang, and Xiqi Gao are with the National Mobile Communications Research Laboratory, Southeast University, Nanjing 210096, China, and also with the Purple Mountain Laboratories, Nanjing 211100, China (e-mail: xiaoxiang@seu.edu.cn; lyou@seu.edu.cn; wangwj@seu.edu.cn; xqgao@seu.edu.cn).
		
		%Jue Wang is with School of Information Science and Technology, Nantong
		%University, Nantong 226019, China, and also with Nantong Research Institute
		%for Advanced Communication Technologies, Nantong 226019, China (e-mail: wangjue@ntu.edu.cn).
		%}
}

%\markboth{Journal of \LaTeX\ Class Files,~Vol.~18, No.~9, September~2023}%
%{Viewing Channel Knowledge as Pixel: An I2I Inpainting Approach for Efficient Channel Knowledge Map Construction}

\maketitle

\begin{abstract}
Channel knowledge map (CKM) has received widespread attention as an emerging enabling technology for environment-aware wireless communications. It involves the construction of databases containing location-specific channel knowledge, which are then leveraged to facilitate channel state information (CSI) acquisition and transceiver design. In this context, a fundamental challenge lies in efficiently constructing the CKM based on a given wireless propagation environment. Most existing methods are based on stochastic modeling and sequence prediction, which do not fully exploit the inherent physical characteristics of the propagation environment, resulting in low accuracy and high computational complexity. To address these limitations, we propose a Laplacian pyramid (LP)-based CKM construction scheme to predict the channel knowledge at arbitrary locations in a targeted area. Specifically, we first view the channel knowledge as a 2-D image and transform the CKM construction problem into an image-to-image (I2I) inpainting task, which predicts the channel knowledge at a specific location by recovering the corresponding pixel value in the image matrix. Then, inspired by the reversible and closed-form structure of the LP, we show its natural suitability for our task in designing a fast I2I mapping network. For different frequency components of LP decomposition, we design tailored networks accordingly. Besides, to encode the global structural information of the propagation environment, we introduce self-attention and cross-covariance attention mechanisms in different layers, respectively. Finally, experimental results demonstrate that the proposed scheme outperforms the benchmark, achieving higher reconstruction accuracy while with lower computational complexity. Moreover, the proposed approach has a strong generalization ability and can be implemented in different wireless communication scenarios.
\end{abstract}

\begin{IEEEkeywords}
	Integrated sensing and communications, digital twin, environment-aware wireless communications, channel knowledge map, channel gain map, laplacian pyramid, attention mechanism.
	%Integrated sensing and communications network, environment-aware wireless communications, channel knowledge prediction, channel path gain maps, Laplacian pyramid, attention mechanism, machine learning
	%integrated sensing and communications, non-geostationary satellite, LEO satellite, massive MIMO, hybrid precoding, beam squint effects, statistical CSI, energy efficiency.
\end{IEEEkeywords}
%\newpage
\section{Introduction}\label{sec:net_intro}
\IEEEPARstart{T}{he} deep integration of mobile communication technology and artificial intelligence will drive the evolution of the fifth generation (5G) and beyond the fifth generation (B5G) to the sixth generation (6G) at both the technical and business levels. 6G will push society towards ``digital twin'' and ``smart ubiquitous'', realizing the integration and interaction between the physical and the virtual world. Emerging applications, such as indoor localization and Wi-Fi sensing, etc., require higher end-to-end information processing capability in 6G networks. In addition to pursuing extremely low latency and other ultra-high performance indicators, it will also build an integrated sensing and communications (ISAC) network \cite{9737357,10054381,you,9793704}.
%As the standardization of 5G solidifies, 信息技术、移动通信技术、人工智能与大数据技术的深度融合（ICDT），驱动着5G 在技术和业务两个层面向6G 演进，业务要素从人向智能体、物理空间和虚拟空间要素扩展，信息处理功能需求从信息传递向信息采集、信息计算扩展。面向2030年，第六代移动通信将推动社会走向“数字孪生”和“智慧泛在”，实现物理世界与虚拟世界的融合和交互。新兴业务对6G网络提出了更高的端到端的信息处理能力，使其除了追求super high capacity, extremely low latency, and ultra-massive connectivity,ultra-high positioning``'' accuracy等超高性能指标之外，还将打造integrated sensing and communications (ISAC) network.

With the rise in the number and density of connected devices, the expansion of antenna array dimensions, and the wider bandwidth usage, ISAC networks are expected to involve ultra-large dimensional wireless channels. Conventional reliance on pilot-based channel training and feedback methods to acquire real-time CSI may have prohibitively high overhead \cite{9373011}. Note that the propagation environment, such as the geometric location relationship in city or terrain maps, is not only static, but also a key factor that affects the parameters of the channel and the performance of a wireless communication system. As such, for ISAC networks, environment-aware wireless communications have attracted significant research interest and attention in both academia and industry \cite{zhuyongxiang}.

CKM plays a vital role as a bridge in enabling environment-aware wireless communications, which provides location-specific channel knowledge associated with a potential base station (BS) to any (B2X) pairs. Specifically, CKM, also referred to as channel fingerprint, functions as a site-specific database tagged with the precise locations of both transmitters and receivers. Within this database, essential channel-related details are stored, providing valuable information such as channel gain, shadowing, angle of arrival/departure, and channel impulse responses, etc. These details are instrumental in alleviating the longstanding challenge of high complexity in CSI acquisition by effectively utilizing propagation environment information. According to the different attributes of the channel contained in CKM, it is divided into different categories, such as channel gain map (CGM), channel shadowing map, and beam index map \cite{9373011}. One typical category of CKM is CGM, which is used to predict channel gains at specific locations in a targeted area.

Traditional channel modeling methods often utilize stochastic modeling (SM) \cite{nurmela2015deliverable}, which relies on some specific assumptions and probability distributions of channel parameters or components. Unfortunately, due to the influence of the propagation environment, these harsh preconditions are often not satisfied in the actual wireless communication. In contrast, environment-aware enabled CKM is through learning of the physical propagation environment, while directly reflecting the intrinsic propagation environment features, does not rely on any assumption on additional channel parameters.
%根据CKM所包含的传播信道不同的属性，其可被进一步细分。传统的信道建模方式往往采用随机建模的方式，它是建立在信道参数或成分的某些特定的假设和概率分布上，不幸的是，受无线传播物理环境的影响，这些苛刻的前提条件往往在实际无线通信过程中不能被一一满足。相反，环境感知使能的CKM是通过对无线传播物理环境的学习，而直接反应的内在的无线传播环境，不依赖于额外信道参数的假设

%The existing CKM works mainly focus on terrestrial networks and have explored many promising applications in the space-air-ground-sea integrated network. 
CKM-enabled environment-awareness communications have been recently studied for various applications in the space-air-ground integrated network. For example, CKM is especially appealing for future wireless communication systems with massive multiple-input multiple-output (MIMO) and prohibitive channel training overhead \cite{9617121}. Specifically, in the context of massive MIMO communication, beam alignment techniques based on pilot training or beam sweeping will bring unaffordable overhead with the rapid increase in antenna dimensions and the number of user devices. To this end, \cite{9473871,9373011,9681979} employ the CKM at the BS, together with the user location information, to acquire critical channel parameters, allowing the MIMO channel matrix to be reconstructed without relying on traditional channel training methods. 
	
Another application scenario for CKM-assisted communications involves non-cooperative nodes, including interferers and eavesdroppers. The design of cognitive and secrecy communications necessitates the acquisition of CSI relating to both primary users and eavesdroppers. However, it presents a challenge due to their non-cooperative nature. Assisted by CKM, it becomes feasible to directly infer the CSI from the location data of non-cooperative nodes \cite{zeng2024tutorial}. Additionally, in the ISAC scenario, CKM continues to hold significance. Once the CKM of the target area is developed, the one-to-one mapping relationship between location and channel-related parameters is established. Consequently, the location of a user equipment can be predicted based on its corresponding channel-related parameters \cite{yang2013rssi}. Moreover, the geometric distribution of channel features can serve to infer the locations and shapes of obstacles, such as constructing obstacle maps from path loss measurements \cite{zhang2020constructing, 9838964}. 
	
More recently, CKM has garnered increased attention as an enabler for cellular-connected unmanned aerial vehicle (UAV) communications. By leveraging CKM, these UAVs can adapt and optimize their communication strategies in real time, ensuring efficient and reliable communication during their missions \cite{9814544,9269485}. In addition, CKM can also be applied to other typical scenarios, such as device-device (D2D) sub-band assignment \cite{9373011}, user-cell site association \cite{7314981}, power control in massive MIMO systems \cite{9104036, 6697931}, activity detection for massive connectivity \cite{8264818}. From the aforementioned studies, it is evident that the CKM plays a crucial role in enabling environment-aware wireless communication.

However, one of the most fundamental and critical issues for CKM-enabled environment-aware communication lies in efficiently constructing CKM based on the geometric location relationship of city maps or the limited measurements at specific locations. The more general approaches are based on data-driven interpolation, which obtains the channel gains at non-measured locations by some interpolation techniques, such as Kriging \cite{9316892}, tensor completion \cite{9523765}, the inverse distance weighted (IDW) \cite{9373011}, nearest neighbours (NN) \cite{9681979}, etc., under the assumption that the channel gains at some locations have been measured. In addition, some papers study algorithms based on model-driven parameter fitting. They combine the given channel gains at specific locations with a priori assumptions of the wireless propagation model to estimate the channel gains corresponding to non-measured locations \cite{8662745,8272409}. Some recent works have adopted algorithms based on deep learning, such as \cite{8865025}, \cite{9373011} using fully connected feedforward neural networks, which only use simple location information, such as the 2D location coordinates, height of the B2X pair or the geometric distance between them. However, except \cite{9354041,9500970,9771088}, the majority of existing CKM construction methods based on interpolation, model parameter fitting and deep learning do not fully consider the specific wireless propagation environment. Therefore, there are several limitations in the implementation process, such as too many location points to be measured, huge storage capacity, high computational complexity, and low estimation accuracy.

In order to efficiently construct CKM, unlike the existing methodologies, we propose an end-to-end LP-based CKM construction scheme, which transforms the problem of CKM construction into an image-to-image (I2I) inpainting task. Specifically, the main contributions of the paper are summarized as follows:
\begin{itemize}
	\item Inspired by \cite{9354041,9500970,9771088}, we explore the intrinsic similarity between the channel knowledge estimation and I2I inpainting tasks. Capitalizing on the similarity between the two tasks, we view the channel knowledge as a 2-D image to design the learning framework, which simultaneously estimates the channel knowledge at each location in a targeted area by inpainting the corresponding pixel in the image matrix.
	\item 
	Conventional I2I inpainting approaches rely on autoencoder or Unet architectures, which usually need to deploy a parameterized encoding and decoding framework to disentangle the input's content and attributes in the low-dimensional latent space, and subsequently reconstruct them back into the original input's dimensionality. Such a process involves expensive computational costs and is hard to scale to higher-resolution tasks. Additionally, in autoencoder architectures, the geometric location information of maps is lost in the irreversible down- and up-sampling processes, leading to degradation of generated results. Meanwhile, Unet architectures, despite preserving some details via skip connections, face issues with increased memory consumption and computational load. Inspired by the closed-form frequency band decomposition and reversible reconstruction framework of an LP \cite{1095851}, we employ parameter-free LP to replace traditional encoding and decoding frameworks, operating in a closed form to enhance the efficiency of image disentanglement and reconstruction, while avoiding the loss of detail information.
    \item Our work focuses on a widely adopted dataset \cite{9354041}. Upon analyzing this dataset, we discover an intriguing property: Based on the LP decomposition, the difference between the geometric location map and the desired CGM is more pronounced in low-frequency components than in high-frequency components. Capitalizing on this property, we design dedicated subnetworks for both low-frequency and high-frequency components. These subnetworks are collaboratively employed to efficiently complete the CGM reconstruction.
    \item To achieve a lightweight model while enhancing the interaction of global and local information, we propose three key modules, including Lightweight Residual Dilated Convolutions (LRDC), Multi-Head Self-Attention (MHSA) and Multi-Head Cross-Covariance Attention (MHCCA). Experimental results show that the proposed method is competitive with the benchmarks regarding computational complexity and reconstruction accuracy in various scenarios.
\end{itemize}

The remainder of this paper is organized as follows. We introduce the system model, and formulate the CGM construction problem in Section II. In Section III, we elaborate on the proposed LP-based CGM construction scheme, including transforming the CGM construction problem to an I2I inpainting task, interesting discoveries on the LP decomposition, and the design of modules and networks. Section IV presents a comprehensive performance evaluation of the proposed model. Finally, conclusions are drawn in Section V.
\section{System Model And Problem Formulation}\label{sec:sys_mod}
In this section, we first introduce the system model, including the propagation environment in a targeted area, the corresponding signal model, as well as the channel gain which is to be estimated in CGM construction. Then, we state the CGM construction problem and view it from two perspectives: Conventional wireless communication and computer vision (I2I inpainting), respectively.
%Then, we further define the channel path gain based on the established channel model, and specifically formulate the CGM estimation problem and its challenges. Finally, we propose to view this problem from a new perspective, that is, address the CGM estimation problem by an image-to-image restoration method, and the objective function to be optimized for the transformed problem is presented.
%Then,  我们基于所建立的信道模型进一步定义了信道路径增益，并具体阐述了CGM估计问题。最后，我们从一种新的角度来看这个问题，通过一种图像到图像的恢复方法来解决CGM的估计问题，并给出了转换之后的新问题所对应需要优化的目标函数。
\subsection{System Model}\label{subsec:chmod}
\begin{figure}[!t]
	\centering
	\includegraphics[scale=0.167]{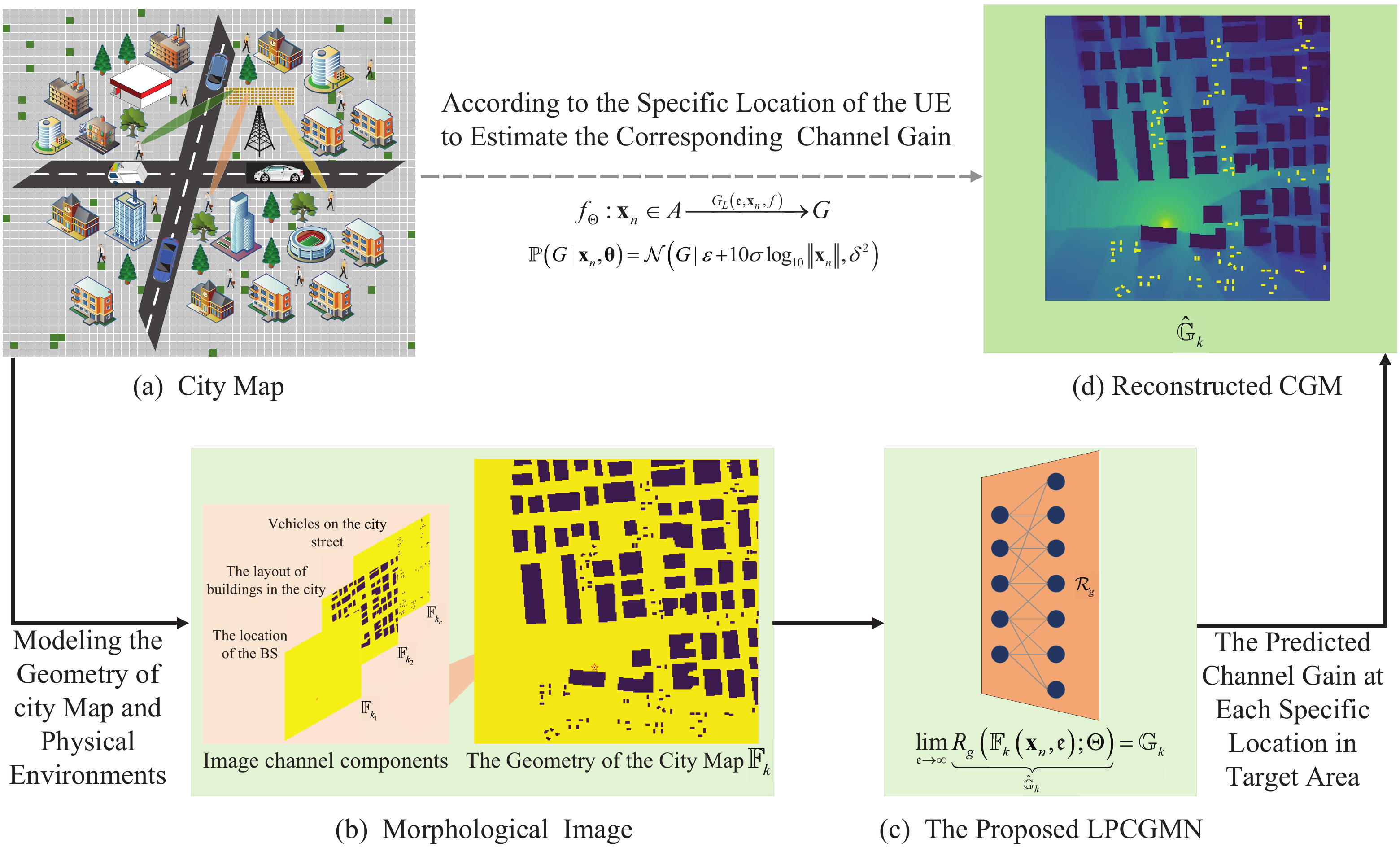}
	\captionsetup{font=footnotesize}
	\caption{The CGM estimation problem is transformed into an I2I inpainting task. We aim to construct the CGM (d) from the wireless environment (a), including buildings, vehicles, etc. The gray dotted arrow represents conventional methods, and the black solid arrow is our method. (b) is the morphological form of the city map. The red pentagram is the location of the BS, these large polygons are buildings, these small yellow rectangles are vehicles, and the yellow shadow is the predicted channel gain at each location. }
	\label{fig:transf}
\end{figure}

Consider a wireless communication scenario in a square coverage area $A \subset {\mathbb{R}^2}$ as shown in \figref{fig:transf} (a), consisting of a base station (BS) and $N$ user equipments (UEs). The signal power attenuation, observed at a UE, may be caused by different factors, such as propagation losses of different paths, reflections and diffractions from buildings, waveguide effects in streets, and obstacle blockages \cite{9354041}. \textit{Among these effects, the relatively slow-varying parts together contribute to the channel gain function, denoted as ${G_L}\left( \mathfrak{e},\mathbf{x}_n,f\right)$ hereafter which describes the large-scale signal attenuation measured at UE locations $\left\{ {{{{\mathbf{x}}_n}}} \right\}_{n = 1}^N = {A}$, and at a finite set of frequencies $f \in \{ {f_1},{f_2},...,{f_{{N_f}}}\} $.}
%\textit{The channel path gain function is a map that obtains to each BS-UE pair of locations $x$, $y$ the corresponding large-scale signal attenuation ${G_L}(x,y)$.} 
%Besides considering large-scale fading effect, the construction of the CGM is also affected by small-scale fading, due to the superposition of scattered wavefronts with different phases at the UE location \cite{9354041}. 
It is worth noting that ${G_L}\left( \mathfrak{e},\mathbf{x}_n,f\right)$ is significantly influenced by the propagation environment, denoted as $\mathfrak{e}$. Besides, the small-scale effects are typically modeled as a complex Gaussian random variable $H$ with unit variance. The received baseband signal at UE is described as
\begin{align}\label{eq:chmd}
	Y = \sqrt {{{G_L}\left(\mathfrak{e},\mathbf{x}_n,f\right) }}HX + Z
\end{align}
where $X$ is the transmitted signal symbol with power ${P_{\cal X}}$, and $Z$ is the additive noise with single sided power spectral density ${N_0}$. The average received energy per symbol is
\begin{align}\label{eq:chmd}
	{{P_{\cal Y}}}=\frac{{\mathbb{E}[{{\left| Y \right|}^2}]}}{B} = \frac{{{{G_L}\left( \mathfrak{e},\mathbf{x}_n,f\right) }{{P_{\cal X}}}}}{B} + {N_0},
\end{align}
where $B$ denotes the signal bandwidth, and the Signal to Noise Ratio (SNR) at the input of the UE baseband processor is, therefore, % MathType!MTEF!2!1!+-
% feaahqart1ev3aaatCvAUfeBSjuyZL2yd9gzLbvyNv2CaerbuLwBLn
% hiov2DGi1BTfMBaeXatLxBI9gBaerbd9wDYLwzYbItLDharqqtubsr
% 4rNCHbWexLMBbXgBd9gzLbvyNv2CaeHbl7mZLdGeaGqiVu0Je9sqqr
% pepC0xbbL8F4rqqrFfpeea0xe9Lq-Jc9vqaqpepm0xbba9pwe9Q8fs
% 0-yqaqpepae9pg0FirpepeKkFr0xfr-xfr-xb9adbaqaaeGaciGaai
% aabeqaamaabaabauaakeaaqaaaaaaaaaWdbiaadofacaWGobGaamOu
% a8aacqGH9aqpdaWcgaqaaiaadEeadaWgaaWcbaGaamitaaqabaGcca
% GGOaGaamiEaiaacYcacaWG5bGaaiykaiaadcfadaWgaaWcbaGaamiw
% aaqabaaakeaacaWGobWaaSbaaSqaaiaaicdaaeqaaOGaamOqaaaaaa
% a!4D34!
${\rm SNR} = {{{G_L}\left( \mathfrak{e},\mathbf{x}_n,f\right) {P_{\cal X}}} \mathord{\left/
		{\vphantom {{{G_L}\left( x,y\right) {P_{\cal X}}} {{N_0}B}}} \right.
		\kern-\nulldelimiterspace} {{N_0}B}}$. 
%Following similar approaches in \cite{7051266}, we translate the standard channel model (1) into an equivalent normalized form that is more conducive for the study of practical signal propagation. Therefore, Eq. (1) can be further rewritten, and the normalized received signal at the $n$th UE is described as
%\begin{align}\label{eq:chmd}
%	{Y_n}(t) = \sum\limits_{i = 1}^L {\sqrt {{{\cal P}^{{\beta _{ni}}}}} } {e^{j{\theta _{ni}}}}{X_i}(t,f) + {Z_n}(t,f),
%\end{align}
%where $\cal P$ is a defined scale parameter, $L$ is the number of wireless propagation paths, ${\sqrt {{{\cal P}^{{\beta _{ni}}}}} }$ and ${e^{j{\theta _{ni}}}}$ are the magnitude and the phase, respectively, of the channel between the BS and the $n$th UE. ${{\beta _{ni}}}$ is the channel strength level of the link between the BS and the $n$th UE on the $i$th path, which can be defined by ${\beta _{ni}} = \log {\rm{SN}}{{\rm{R}}_{ni}}/\log {\cal P}$. 
Inspired by \cite{geng2015optimality} and \cite{9354041}, the channel gain in dB scale is defined as
\begin{align}\label{eq:chmd}
	{G}\left(\mathfrak{e},\mathbf{x}_n,f\right) = {\left( {{P_{\cal Y}}}\right) _{\rm{dB}}} - {\left( {P_{\cal X}}\right) _{\rm{dB}}},
\end{align}
%${P_{\cal Y}}$
which describes the receiver power variation (induced by both the propagation and noise environment) at location $\mathbf{x}_n$. Accurately estimating ${G}$ at each location is the key to constructing an fine-grained CGM. Truncating and rescaling the channel gain function in dB scale is necessary to adapt it for the deep learning-based prediction method proposed subsequently.

For any $\mathbf{x}_n \in {A}$, our goal is to predict the corresponding $G$ defined in (3). To this end, a B2X CGM ${f_\Theta }$ needs to be constructed, which provides mapping from all possible user locations $\mathbf{x}_n$ to the corresponding ${G}$, i.e.,
\begin{align}\label{eq:chmd}
	{f_\Theta }:{{\mathbf{x}}_{{n}}} \in {A}\xrightarrow{{{G_L}\left( {\mathfrak{e},{{\mathbf{x}}_{{n}}},f} \right)}}{G}. 
\end{align} 
Building this mapping (4) is tricky because ${G_L}\left( \mathfrak{e},\mathbf{x}_n,f\right)$ is implicitly affected by the propagation environment, for which analytical path loss and shadowing models can only provide a rough approximation. Therefore, it is challenging to accurately acquire the global and completed CGM corresponding to the targeted area.
%Furthermore, the GIWSN is modeled as a connectivity graph $\textbf{G}(\textbf{V},\textbf{E})$, in which these UE are defined as $\textbf{V} = \{ 1,2,...,N\}$ and the channel link path are denoted by $\textbf{E} \in {\{ {E_{i,j}}\} _{N \times N}}(i,j \in \textbf{V})$. The spatial location of UE node j is denoted by $E{U_j} = ({x_j},{y_j})$. Due to the constraints on transmit power, we assume that the transmission range of each node is limited to a distance ${R_{\max }}$. As a consequence, the channel path ${E_{i,j}}$ is established when the Euclidean distance between nodes $i$ and $j$, denoted by ${D_{i,j}}$, is less than ${R_{\max }}$. 
%Furthermore, the GIWSN is modeled as a connectivity graph $\textbf{G}(\textbf{V},\textbf{E})$, in which these Rx are defined as $\textbf{V} = \{ 1,2,...,N\}$ and the channel link path are denoted by $\textbf{E} \in {\{ {E_{i,j}}\} _{N \times N}}(i,j \in \textbf{V})$. The spatial location of Rx node j is denoted by ${\chi _j} = ({x_j},{y_j})$. Due to the constraints on transmit power, we assume that the transmission range of each node is limited to a distance ${R_{\max }}$. As a consequence, the channel path ${E_{i,j}}$ is established when the Euclidean distance between nodes $i$ and $j$, denoted by ${D_{i,j}}$, is less than ${R_{\max }}$.
\subsection{Conventional Methods vs. Our Perspective}
%\subsubsection{CGM Construction Problem}
%Assuming that a wireless communication system has been deployed in the considered area $A \in {\mathbb{R}^2}$, with a BS and some UEs, whose potential locations $\mathbf{x_n}$ are denoted by the set $Q$. For any given $\mathbf{x_n} \in Q$, our goal is to predict the interested location-specific channel knowledge, denoted as $\nu$, as accurately as possible before real-time channel training is applied. Note that for a more general formulation, the channel knowledge $\nu$ can be any useful information related to the wireless channel, such as the channel gain, shadowing, angle of arrival/departure (AoA/AoD), or even the channel impulse response. In this paper, we focus on predicting the channel gain, previously defined in (4). To this end, a BS-to-any (B2X) CGM ${f_\Theta }$ is constructed, which provides mapping from each UE location $\mathbf{x_n}$ to the corresponding channel gain ${P_g}$, i.e., ${f_\Theta }:\mathbf{x_n} \in Q \to {P_g}$. Due to the limited number of the measured locations $\mathbf{x_n}$ (and the corresponding ${p_g}$ measured at these locations), it is challenging to accurately acquire the global and completed CGM with such a limited environmental sampling database.
%we can receive channel path gain data from the sensor nodes and store them in the edge computing server. 
%\subsubsection{Review of Conventional Wireless Communication}
\subsubsection{Review of Conventional Methods}
Conventionally, stochastic modeling \cite{9373011,nurmela2015deliverable} is used for location-based channel gain estimation, which relies on firstly establishing an analytical model framework, then describing the stochastic distributions of related channel parameters. Specifically, denote the channel gain as ${\tilde G}$, and the required model parameters, arranged in a vector, as $\boldsymbol{\theta }$, the probability density function of ${\tilde G}$ for UE at $\mathbf{x}_n$ is defined as $\mathbb{P}\left( {{\tilde G}|\mathbf{x}_n,\boldsymbol{\theta}} \right)$. According to the classic path loss model, ${\tilde G}$ is denoted as
\begin{align}\label{eq:chmd}
	{\tilde G} = \varepsilon  + 10\sigma  {\log _{10}}\left\| \mathbf{x}_n \right\| + S,
\end{align}
where $\sigma$ and $\varepsilon$ represent the path loss exponent and intercept, respectively. $S \sim {\cal N}\left( 0,{\delta ^2}\right) $ captures the log-normal shadowing with variance $\delta ^2$, and $\left\|  \cdot  \right\|$ denotes the Euclidean norm. In this case, we have $\boldsymbol{\theta}  = [\varepsilon ,\sigma ,{\delta ^2}]$, and the conditional PDF of ${\tilde G}$ is
\begin{align}\label{eq:chmd}
	\mathbb{P}&\left( {{\tilde G}|\mathbf{x}_n,\boldsymbol{\theta} } \right) = \mathcal{N}\left( {{\tilde G}|\varepsilon  + 10\sigma {{\log }_{10}}\left\| {\mathbf{x}_n} \right\|,{\delta ^2}} \right)\nonumber \\
	&= \frac{1}{{\sqrt {2\pi {\delta ^2}} }}\exp \left( { - {{{{\left( {{\tilde G} - \varepsilon  - 10\sigma {{\log }_{10}}\left\| {\mathbf{x}_n} \right\|} \right)}^2}} \mathord{\left/
				{\vphantom {{{{\left( {{G} - \varepsilon  - 10\sigma {{\log }_{10}}\left\| {\left( x,y\right) } \right\|} \right)}^2}} {2{\delta ^2}}}} \right.
				\kern-\nulldelimiterspace} {2{\delta ^2}}}} \right),
\end{align}
where $\mathcal{N}\left( {\tilde G}|\mu ,{\delta ^2}\right) $ is the probability density function of a Gaussian random variable ${\tilde G}$ with variance ${\delta ^2}$ and mean $\mu $. 

Estimating channel knowledge using the methods described above may have limitations in performance, since it usually does not consider a priori environmental information when determining the model structure, except for some general attributes like environmental types, e.g., urban, suburban\cite{9373011}. For example, under the channel modeling of (5), the predicted channel gains at specific locations around the BS appear symmetric regardless of the accuracy of the chosen parameter vector $\boldsymbol{\theta }$. However, this symmetric prediction contrasts with reality, where blockage and scattering in different directions are typically asymmetric.
%since it does not make full use of the available prior knowledge about the actually complicated radio environment. 
%Moreover, if an interpolation-based algorithm is used to estimate the channel gain, the computational complexity will show a cubic growth in the ratio of the target region size and the interpolation resolution \cite{deng2020radio}. 
%\subsubsection{View the Problem from the Perspective of Computer Vision}
%\subsubsection{A Fresh View Look at CGM}
\subsubsection{A Computer Vision Perspective Look at CGM}
Stochastic modeling ignores the environmental information, such as the morphology of streets, buildings, and spatial geometric relationships. In contrast, in the typical tasks of I2I inpainting, the focus of feature extraction and representation is to learn the intrinsic structure of the image. I2I inpainting, as the name suggests, is to recover the damaged part of the image. This inspires us to apply I2I inpainting techniques in the CGM construction problem. \textit{Let the input be a geometric location map and the output be the desired CGM, the difference between the two is the channel gain to be estimated, which can actually be viewed as the content to be repaired in I2I inpainting task.} As shown in \figref{fig:transf}, we first transform the CGM construction problem into an I2I inpainting task, which predicts the channel gains at UE locations by recovering the corresponding pixel values in the image matrix. Then we propose an efficient reconstruction network ${\mathcal{R}_g}$ for CGM construction and let $\Theta $ represent the set of parameters to be learned in ${\mathcal{R}_g}$. In supervised training mode, a training set of input city maps ${\mathbb{F}_k}\left( {\mathbf{x}_n,\mathfrak{e}} \right)$ and output corresponding CGM ${\mathbb{G}_k}$ are given, where $k=1,...,K$ and $K$ is the number of training samples. City map ${\mathbb{F}_k}\left( {\mathbf{x}_n,\mathfrak{e}} \right)$ mainly contains environmental features $\mathfrak{e}$, such as BS location, buildings, roads, vehicles, which are saved in PNG image format.
\textit{The goal of model training is to continuously optimize the parameter set $\Theta $ of ${\mathcal{R}_g}$, using ${\mathbb{F}_k}\left( {\mathbf{x}_n,\mathfrak{e}} \right)$ as the prior knowledge, to efficiently reconstruct the desired CGM, i.e.,}
\begin{align}\label{eq:chmd}
	\mathop {\lim }\limits_{\mathfrak{e} \to \infty } \underbrace {{{\cal R}_g}\left( {{{\mathbb{F}_k}}\left( {{{\bf{x}}_n},\mathfrak{e}} \right);\Theta } \right)}_{\hat{\mathbb{G}}_k} = {\mathbb{G}_k},
\end{align}
	 %${\mathbb{G}_k} \approx {\mathcal{R}_g}\left( {{\mathbb{F}_k}\left( {x,y} \right);\theta '} \right)$ for every $k = 1,2,...,K$
	%$\mathop {\lim }\limits_{\mathfrak{e} \to \infty } \underbrace {{{\cal R}_g}\left( {{{\mathbb{F}_k}}\left( {{{\bf{x}}_n},\mathfrak{e}} \right);\Theta } \right)}_{\hat{\mathbb{G}}_k} = {\mathbb{G}_k}$ 
	%by using ${\mathbb{F}_k}\left( {\mathbf{x}_n,\mathfrak{e}} \right)$ as the prior knowledge, 
	%where ${\mathbb{F}_k}\left( {\mathbf{x}_n,\mathfrak{e}} \right)$ is the $k$-th city map sample with the wireless propagation environment $\mathfrak{e}$ and BS location, ${\mathbb{G}_k}$, 
where ${\hat{\mathbb{G}}_k}$ is the reconstructed CGM. Therefore, from the perspective of I2I inpainting, the CGM construction problem can be formulated as
%The above optimization process is usually done by backpropagation of the gradient, so we optimize the following objective function in the direction of gradient descent

%\begin{align}\label{eq:chmd}
%	\mathop {{\text{minimize}}}\limits_{\{ \theta '\} } {\text{   %}}\mathcal{L}(\theta ') = \mathbb{E}\left\{ {\sum\limits_k {\left\| %{{\mathbb{G}_k} - {\mathcal{R}_g}({\mathbb{F}_k};\theta ')} \right\|} } %\right\}.
%\end{align}
%\[\left\| x \right\|_2^2\]
\begin{subequations}\label{eq:spdb}
	\begin{align}
		\mathop {{\rm{min}}}\limits_{{\Theta }} \mathcal{L}( {\mathbb{G}_k},&{\hat{\mathbb{G}}_k})   = \mathbb{E}\left\{ {\left\| {{\mathbb{G}_k} - {\mathcal{R}_g}({\mathbb{F}_k}({{\mathbf{x}}_n},\mathfrak{e});\Theta )} \right\|_2^2} \right\},\label{eq:dbso}\\
		%\rm{s.t.}\quad &{\hat{\mathbb{G}}_k} = {\mathcal{R}_g}( {{\mathbb{F}_k}\left( {\mathbf{x}_n,\mathfrak{e}} \right);\Theta } ),\label{eq:dbso}\\
		&{\rm{s.t.}}\quad {\mathbf{x}_n} \in {A} ,k \in \{ K\},\label{eq:dbsc}
	\end{align}
\end{subequations}
where ${\left\|  \cdot  \right\|_2}$ represents the Euclidean norm. The parameter set $\Theta $ can be learned by gradient descent-based optimizers, such as the adaptive momentum estimation (Adam) optimizer \cite{10061451}.
%\[\mathop {{\text{minimize}}}\limits_{\left\{ {\theta '} \right\}}   \mathcal{L}\left( {{\mathbb{G}_k},{{\hat \mathbb{G}}_k}} \right) = \mathbb{E}\left\{ {\sum\limits_k {\left\| {{\mathbb{G}_k} - {\mathcal{R}_g}\left( {{\mathbb{F}_k}\left( {\left( {x,y} \right),\mathfrak{e}} \right);\theta '} \right)} \right\|} } \right\}\]
%\subsection{Artificial Neural Networks and Image Processing}
%\subsection{High-resolution Image-Image Translation}
%\begin{figure}[!t]
%	\centering
%	%\includegraphics[width=0.8\textwidth]{figure/nc (2).eps}
%	\includegraphics[scale=0.16]{figure/111cm.eps}
%	\captionsetup{font=footnotesize}
%	\caption{The CGM estimation problem is transformed into an I2I inpainting task. We aim to construct the CGM (d) from the wireless environment (a), including buildings, vehicles, etc. The gray dotted arrow represents conventional methods, and the black solid arrow is our method. (b) is the morphological form of the city map. The red pentagram is the location of the BS, these large polygons are buildings, these small yellow rectangles are vehicles, and the yellow shadow is the predicted channel gain at each location. }
%	\label{fig:transf}
%\end{figure}
\section{LP-Based CGM Construction}\label{sec:JCS_hp}
%几乎所有现存的信道地图估计问题均采用插值算法，没有从以往经验中进行学习。然而，这是一个很天然的算法去对目标地理区域内的信道参数，如信道路径增益来进行学习，从而通过预测的方式解决这个估计的问题。但是在实际过程中会遇到一下几个问题：1.目标位置区域中用户的数量可能会很大，如果直接在每个位置设置一个估计器显
%多个点的序列估计问题转换为张量补齐问题；量化；I2I inpainting task
%The sequence estimation problem of multiple points is transformed into a tensor completion problem. Quantization; I2I inpainting task

%在之前的章节，我们阐述了CGM的估计问题，并从I2I修补的角度给出了初步的解决方案。这个章节主要围绕着1）如何将CGM估计问题转换为基于深度学习的I2I inpainting task和2）如何设计一个高效的重构网络来应对高分辨率的几何位置地图。就1）而言，我们首先对目标区域进行空间的离散化，将UE的具体位置转化为图像中像素点的坐标，将UE具体位置对应的channel path gain视为对应的像素值。然后对离散后的目标区域的channel path gain进行像素级别的归一化，其原因会在B小节提到。就2）而言，我们首先分析基于I2I inpainting 方法的高分辨率CGM重构的一些限制，whic which is an important motivation for us to redesign the neural network。Next, 我们给出了一些有趣的发现关于LP在分解高分辨率图像的过程中，这些发现对我们在设计后面的重构网络是有指导意义的. Finally, the proposed 基于LP的重构 model for CGM 估计 is presented.
This section mainly involves the following questions: i) How to specifically transform the CGM construction problem into a deep learning-based I2I inpainting task, and ii) How to design an efficient reconstruction network to cope with high-resolution geometric location maps (city maps). For the first question, we first discretize the targeted area in the spatial domain, convert the UE locations into the coordinates of pixel points in the image matrix, and view the channel gain at each UE location as the corresponding pixel value. Then, the channel gain of the discretized target area is normalized at the pixel level. For the second problem, we first analyze the limitation of I2I inpainting approach, and we reveal some findings in the LP decomposition of high-resolution images, which are instructive for us to design the efficient reconstruction network. Finally, the proposed LP-based reconstruction model for CGM construction is presented.

%\subsection{CGM Construction as a I2I Inpainting Task}
\subsection{Spatial Discretization of the Geometric Location Map}
We first grid the geometric location map using spatial discretization, to facilitate feedforward architectures to solve the CGM construction problem \cite{9523765}. Similar discretizations have been applied in \cite{8272409,7147778,6648369}. In more detail, we discretize the continuous geometric location map into a fixed-size image. Each spatial grid is treated as the corresponding pixel point in the 2D image. As such, estimating the channel gain of $N$ position coordinates can be translated into a super-resolution I2I inpainting.
\begin{figure}[!t]
	\centering
	\includegraphics[scale=0.31]{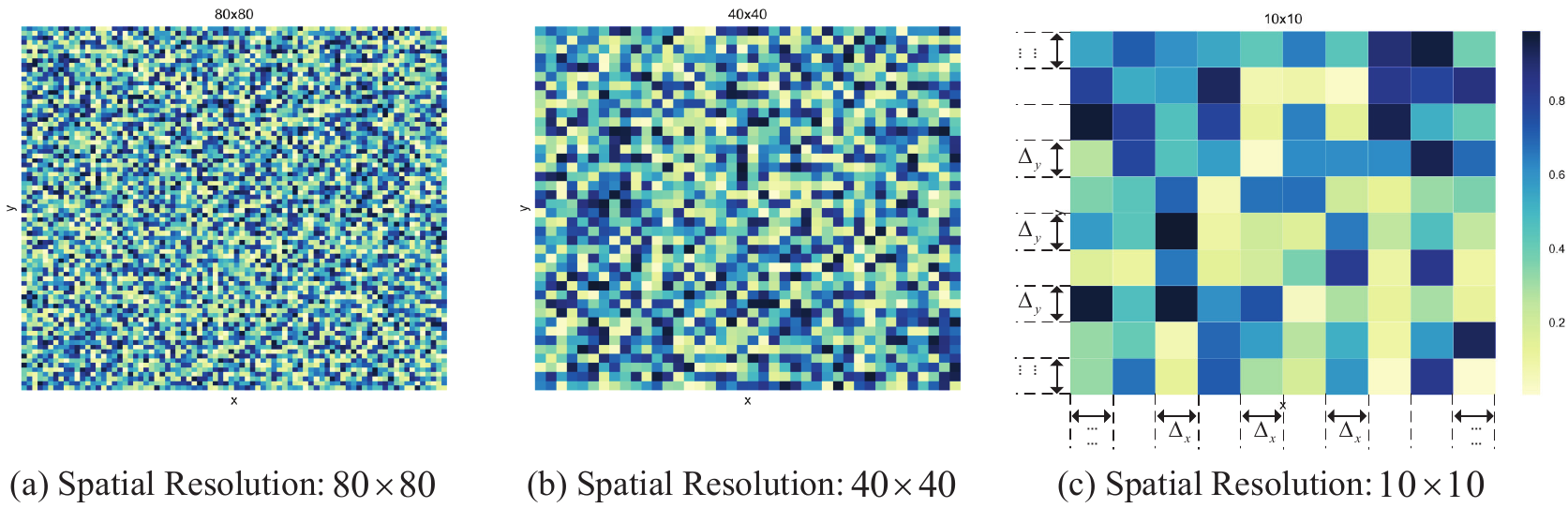}
	\captionsetup{font=footnotesize}
	\caption{The different degrees of spatial discretization corresponding to the geometric location map. (a), (b), and (c) represent the spatial discretization of the target region $A$ into $80 \times 80$, $40 \times 40$, and $10 \times 10$ spatial grids, respectively. Note that ${\Delta _x}$ and ${\Delta _y}$ are taken to be 1, respectively, and the actual geometric location map is gridded into $256 \times 256$ spatial grids in the experimental part of this paper.}
	%In this paper, ${\Delta _x}$ and ${\Delta _y}$ are taken to be 1, respectively, and the geometric location map is gridded into $256 \times 256$ minimal spatial domains}
\label{fig:s}
\end{figure}
Specifically, for the targeted location region $A$, we discretize along its geometric space dimensions, that is, along the X-axis and Y-axis, respectively, taking ${\Delta _x}$ and ${\Delta _y}$ as the minimum interval units, and meshing it into a 2D vector graph with ${N_x}$ rows and ${N_y}$ columns, as shown in \figref{fig:s}. \figrefs{fig:s} (a), (b), and (c) show the representations under different spatial discretization degrees, respectively. Each spatial grid is denoted as ${{\mathbf{\Gamma }}_{i,j}}$, where $i = 1,2,...,N_x$ and $j = 1,2,...,N_y$, the $\left( i, j\right) $-th spatial grid is represented by 
\begin{align}\label{eq:chmd}
{{\mathbf{\Gamma }} _{i,j}}: = {[i{\Delta _x},j{\Delta {}_y]^T}}.
\end{align}
Note that the spatial grid coordinate is the center point of a rectangle. For ease of presentation, let $\mathcal{A}_{i,j}\subset\left\{ {1,...,N} \right\}$ denote the set containing the indices of location coordinates of all UEs in the $(i,j)$-th spatial grid. The process of spatial discretization is according to the criterion of minimum distance, i.e., ${\mathbf{x}_n} \in \mathcal{A}_{i,j}$ iff $\left\| {{{\mathbf{\Gamma }}_{i,j}} - {{\mathbf{x}}_n}} \right\| \leq \left\| {{{\mathbf{\Gamma }}_{i',j'}} - {{\mathbf{x}}_n}} \right\|,\forall (i',j') \ne (i,j)$, where when the equality holds, arbitrarily decide ${\mathbf{x}_n} \in \mathcal{A}_{i,j}$ or ${\mathbf{x}_n}\in \mathcal{A}_{i',j'}$.

By discretizing the geometric locations of UE, the channel gain vector ${G}\left( f\right)  \in {\mathbb{R}^{N \times 1}}$ of the spatial locations $\left\{ {{\mathbf{x}}_n} \right\}_{n = 1}^N$ is rearranged into a 2D channel gain matrix ${\rm \mathbf{G}}\left( f\right)  \in {\mathbb{R}^{{N_x} \times {N_y}}}$. Note that if $N < {N_x}{N_y}$, the component values of vector ${G}\left( f\right)$ do not fully fill the matrix $\mathbf{G}\left( f\right)$ based on the above grid assignment. In this case, 0s are used to fill up the gaps in the matrix $\mathbf{G}\left( f\right)$. The channel gain of the $(i,j)$-th spatial grid is defined as ${[{\rm \mathbf{G}}\left( f\right) ]_{i,j}} = { G}\left( {{\mathbf{\Gamma }}_{i,j}},f\right) $. Further, when considering multiple frequencies, the 2D CGM can be concatenated in frequency dimension to form a tensor representation ${\rm \mathbf{G}} \in {\mathbb{R}^{{N_x} \times {N_y} \times {N_f}}}$, namely
%The wireless frequency band $f \in F$ actually used is usually a finite set, where $F = \{ {f_1},{f_2},...,{f_{{N_f}}}\}$. This frequency dimension can be regarded as the color channel dimension of RGB image in image processing, and the 2D CGM can be concatenated in this dimension to form a tensor representation ${\rm \mathbf{P_g}} \in {\mathbb{R}^{{N_x} \times {N_y} \times {N_f}}}$, namely\[\left| s \right|\]
\begin{align}\label{eq:chmd}
{[{\mathbf{G}}]_{i,j,{n_f}}} = {{G}}\left( {{\mathbf{\Gamma }}_{i,j}},{f_{{n_f}}}\right) .
\end{align}
Clearly, if the discrete geometric location map has a high spatial resolution, i.e., if ${\Delta x}$ and ${\Delta y}$ are small enough, then $\mathop {\lim }\limits_{\Delta x,\,\Delta y \to 0}{{\mathbf{\Gamma }}_{i,j}} = {\mathbf{x}_n}$ holds for all ${\mathbf{x}_n} \in \mathcal{A}_{i,j}$. For $\left|\mathcal{A}_{i,j}\right|>0$ the channel gain of continuous location coordinates can be well approximated, which is expressed as 
\begin{align}\label{eq:chmd}
{G}\left( {{\mathbf{\Gamma }}_{i,j}},f\right)  = \frac{1}{{\left| {\mathcal{A}_{i,j}} \right|}}\sum\nolimits_{{{\mathbf{x}_n}} \in \mathcal{A}_{i,j}} {{G}\left( {{\mathbf{x}}_{n}},f\right) }.
\end{align}
Note that in (11), when $\Delta x$ and $\Delta y$ both go 0, each set $A_{i,j}$ has at most one element. In this case, the channel gain of the spatial grid is equal to that of individual UE. Thus, the summation sign in the right hand side of (11) is just for the notational convenience and there is at most one term in the summation. The transformation of spatial discretization allows for the estimation of channel gains at continuous locations to be converted into the pixel/grid prediction, which also facilitates the design of neural networks in subsequent subsections.

\subsection{Pixel-Level Graying of Channel Gain}
%?????????????2D????????????????????????????
A channel gain ${{G}}\left( {{{\mathbf{\Gamma }}_{i,j}}},f\right) $ corresponding to a spatial grid ${{{\mathbf{\Gamma }}_{i,j}}}$ is treated as the value of the corresponding pixel point in the image matrix, which needs to be grayed for further manipulation. To this end, we first convert its value to the interval range from 0 to 1 using the minimum maximum normalization. The gray-level channel gain is expressed as
\begin{align}\label{eq:chmd}
{{{{G}}}^{\prime }}\left( {{{\mathbf{\Gamma }}_{i,j}}},f \right)=\frac{{{G}}\left( {{{\mathbf{\Gamma }}_{i,j}}},f \right)-\min \left( {{G}}\left( {{{\mathbf{\Gamma }}_{i,j}}},f \right) \right)}{\max \left( {{G}}\left( {{{\mathbf{\Gamma }}_{i,j}}},f \right) \right)-\min \left( {{G}}\left( {{{\mathbf{\Gamma }}_{i,j}}},f \right) \right)},
\end{align}
The importance of gray-level conversion is highlighted in the following two aspects:
\subsubsection{Considerations for Optimizing Neural Network Training}
\begin{figure}[!t]
\centering
\includegraphics[scale=0.32]{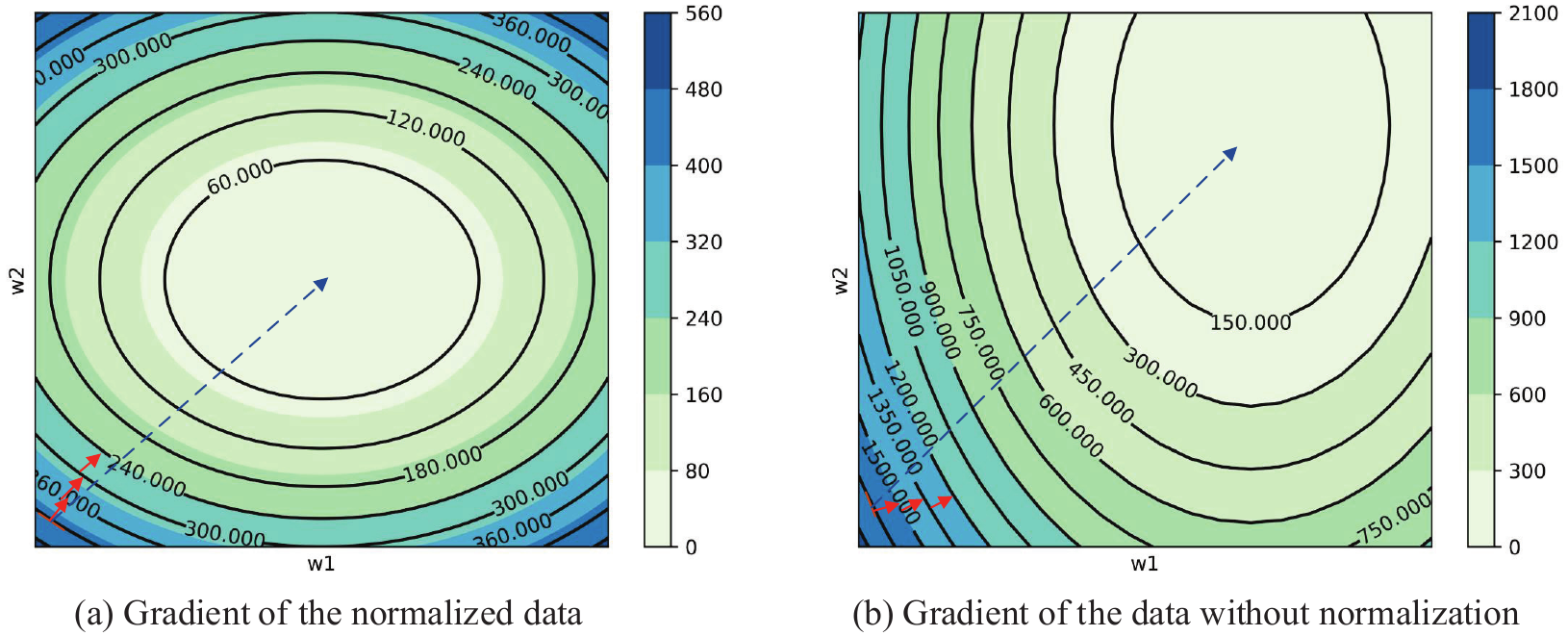}
\captionsetup{font=footnotesize}
\caption{Visual comparison of the gradient descent process for normalized and unnormalized data. The blue and red arrows represent the optimal and actual direction of gradient descent, respectively.}
\label{fig:gradient}
\end{figure}

%A machine learning algorithm is said to have scale invariance if it scales all or part of its features without affecting its learning and prediction.
%Theoretically, neural networks have scale invariance and can adapt to the features of different scales by adjusting the parameters of each network layer. 
Without the gray-level conversion, it is challenging to ensure that all input features with different scales fall within the gradient-sensitive interval of the activation function, and those outside the interval will cause their gradients to disappear, increasing the difficulty of training. When the dimensions of input features are very high, it is trickier to make the activation function work in its gradient-sensitive interval.
Furthermore, the uncertainty introduced by features of varying scales can degrade the efficiency of gradient descent algorithm. Fortunately, as shown in \figref{fig:gradient}, pixel-level normalization can render the optimization landscape of the neural network smoother, stabilize the gradient, and improve the convergence speed. %When we normalized the data to be processed at the pixel level, it is shown in Fig. 3 (a). It can be found that the gradient direction of most positions is approximately consistent with the optimal search direction, which means that the number of gradient descent iterations is less, but also increases the possibility of finding the optimal solution because the direction of each step of the gradient is almost pointing to the minimum value of the objective function, which greatly improves the efficiency of training.

\subsubsection{Considerations for Evaluating CGM Construction Performance} 
When the normalized mean-squared error (NMSE) is used to evaluate the CGM construction, it is given by
\begin{align}\label{eq:chmd}
{\rm NMSE} = \frac{{\sum\limits_{{{\mathbf{\Gamma }}_{i,j}} \in A} {{{\left| {{{\hat {{{G}^{\prime}}}} }\left( {{\mathbf{\Gamma }}_{i,j}},f\right)  - {G}^{\prime} \left( {{\mathbf{\Gamma }}_{i,j}},f\right) } \right|}^2}} }}{{\sum\limits_{{{\mathbf{\Gamma }}_{i,j}} \in A} {{{\left| {{G}^{\prime} \left( {{\mathbf{\Gamma }}_{i,j}},f\right) } \right|}^2}} }},
\end{align}
where $\hat {{{G}^{\prime}}}\left( {{{\mathbf{\Gamma }}_{i,j}}},f\right) $ is the predicted channel gain (after graying) for the $\left( i,j\right) $-th spatial grid. Note that (13) is mainly affected by numerator and denominator components with relatively large absolute values, while the spatial grids with small absolute values of channel gains are negligible in performance evaluation. This is contradictory because for the spatial grid with a large absolute value of channel gain in CGM, the corresponding received signal strength is relatively small, and the contribution to the construction of CGM is also smaller. Therefore, weak signals should not dominate the construction of CGM. This problems can be effectively solved through (12).
%With the help of (12), the channel gain corresponding to the spatial grid with the highest signal strength is converted to 1, indicating that it is essential. The channel gain corresponding to the spatial grid with the lowest signal strength is converted to 0, which means that it is too small to be interesting to our model, which can effectively solve the above problems. 
\subsection{The Laplacian Pyramid for Decomposition of High-Resolution Maps}
\subsubsection{Motivation for Utilizing the Laplacian Pyramid}
Processing high-resolution images is tricky in computer vision \cite{9729564,Liang_2021_CVPR}, mainly due to the trade-off between prediction accuracy and computational complexity, which also exists in our CGM construction task.
%, it is also challenging to provide a high-resolution CGM with affordable computational overhead. To this end, we use Laplacian Pyramid to solve this problem.
%The reason why these two problems are mentioned is that it is also the challenge of image-image inpainting, transformation and other fields in the computer vision \cite{9729564,Yi_2020_CVPR,Yu_2018_CVPR,Kim_2022_CVPR,Liang_2021_CVPR}. 
%Particularly, when a high-resolution geometric location map is input, the number of spatial grids is huge, which means that the neural network model needs to use kernels with larger size and more color-channel dimensions for convolution. 
The traditional I2I method adopts the auto-encoder framework, and converts the high-resolution input into a low-dimensional latent space through convolution downsampling, then reconstructs the target image by deconvolution upsampling the feature representation of low-dimensional latent space. However, these algorithms limit the output image's resolution, or they take a long time in inference models. Moreover, the upsampling and downsampling operations are not completely reversible, which implies that there is inevitable loss of detail information in the process of input-latent space-output \cite{1095851,Liang_2021_CVPR}. 
%In addition, the high-resolution image requires a neural network to spend more attention on exploring the correlation between its content details, which undoubtedly intensifies the computational burden of the model.

\begin{figure*}[!t]
\centering  
\includegraphics[scale=0.177]{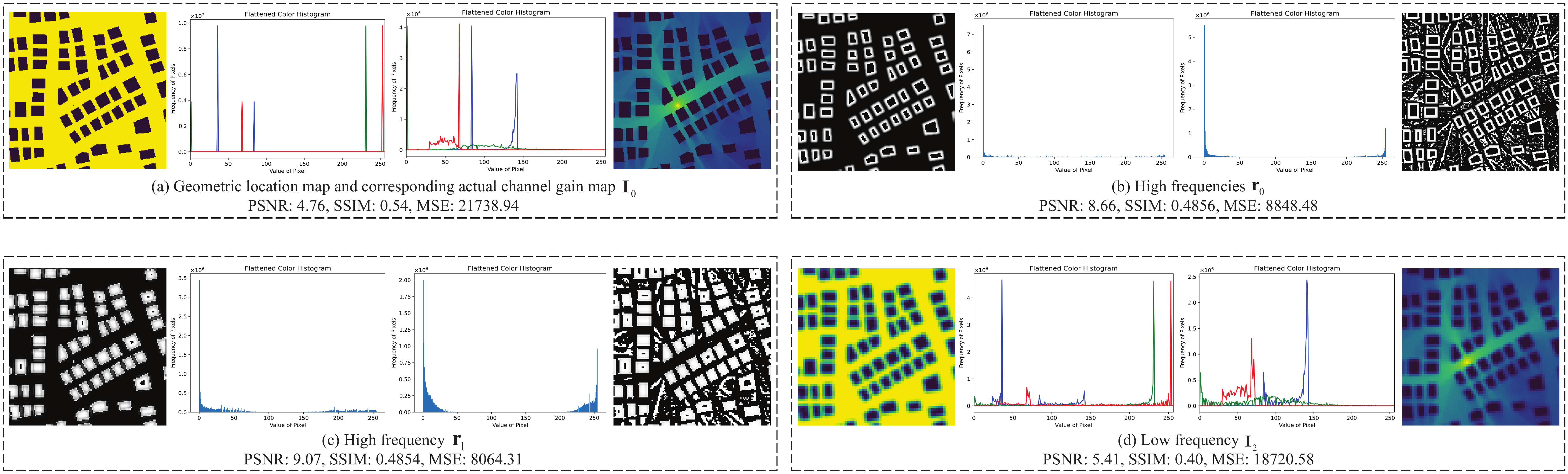}
%\includegraphics[scale=0.177]{figure/lapls_mse_marjor_68.eps}
%\includegraphics[scale=0.177]{figure/lapls_mse_marjor_210.eps}
%\includegraphics[scale=0.177]{figure/lapls_mse.eps}
%\includegraphics[scale=0.177]{figure/lapls_mse_1.eps}
%\includegraphics[scale=0.177]{figure/xinde-lp-mse.eps}
%newnewlapmse.eps
%newnew-lp-mse.eps
%\includegraphics[width=0.3\textwidth]{figure/newlapmse.eps}
\captionsetup{font=footnotesize}
\caption{The LP decomposition obtains the difference between the geometric location map (left) and the CGM (right) for different frequency components. As shown by the MSE, SSIM, PSNR, and the histogram, the differences between the geometric location map and the CGM are more pronounced in the low-frequency components.}
\label{fig:lp}
\end{figure*}
Considering the irreversibility of up-sampling and down-sampling, the LP can retain the difference between each layer of down-sampling and its up-sampling, and the retained residual information can effectively fill in the details lost in image reconstruction \cite{1095851}. In specific, given an input ${I_0}$ with size $h \times w$, it firstly obtains a low-pass estimation ${I_1} \in {\mathbb{R}^{\tfrac{h}{2} \times \tfrac{w}{2}}}$ where each pixel is a weighted average of the neighboring pixels based on an octave Gaussian filter. To ensure the reversibility of the image reconstruction, the LP retains the high-frequency residual information ${r_0}$ as ${r_0} = {I_0} - {\tilde I_0}$, where ${\tilde I_0}$ represents the upsampled result from ${I_1}$. To further decrease the input resolution, the aforementioned operations is performed iteratively by LP on ${I_1}$ to generate a set of low- and high-frequency components. The irreversibility of up-sampling and down-sampling can be alleviated by the decomposition of an LP and the reconstruction of an inverted LP. Inspired by the above, we employ the LP's natural reversible and closed-form decomposition properties to address the aforementioned challenges in the accuracy-complexity trade-off. Furthermore, LP can decompose the high-resolution input into multiple sub-maps with different spatial resolutions, which brings the benefit of fusing multi-scale feature information. 
\subsubsection{Unveiling the Importance of Different Frequency Components}
In image decomposition utilizing the LP, we find some inherent relationships between the geometric location map and its corresponding CGM. For the ease of description, we first define three common performance metrics for comparing differences between a pair of images: Mean Squared Error (MSE), Structural Similarity (SSIM), and Peak Signal-to-Noise Ratio (PSNR), which are respectively defined as ${\rm MSE}=\frac{1}{{{N}_{x}}\times {{N}_{y}}}\sum\nolimits_{i=0}^{{{N}_{x}}}{{{\sum\nolimits_{j=0}^{{{N}_{y}}}{\left[ \left| \mathbf{O}\left( i,j\right) -\mathbf{I}\left( i,j\right)  \right| \right]}}^{2}}}$, ${\rm SSIM}\left( \mathbf{O},\mathbf{I}\right)  = \frac{{\left( 2{u_\mathbf{I}}{u_\mathbf{O}} + {C_1}\right) \left( 2{\delta _{\mathbf{IO}}} + {C_2}\right) }}{{\left( u_\mathbf{I}^2u_\mathbf{O}^2 + {C_1}\right) \left( \delta _\mathbf{I}^2\delta _\mathbf{O}^2 + {C_2}\right) }}$, ${\rm PSNR} = 20{{\rm{log}}_{10}}\left( {\frac{{{2^b} - 1}}{{\sqrt {\rm{MSE}} }}} \right)$,
%\begin{align}\label{eq:chmd}
%  MSE=\frac{1}{{{N}_{x}}\times {{N}_{y}}}\sum\nolimits_{i=0}^{{{N}_{x}}}{{{\sum\nolimits_{j=0}^{{{N}_{y}}}{\left[ \left| O(i,j)-I(i,j) \right| \right]}}^{2}}},
%\end{align}
%\begin{align}\label{eq:chmd}
%	SSIM(\mathbf{O},\mathbf{I}) = \frac{{(2{u_I}{u_O} + {C_1})(2{\delta _{IO}} + {C_2})}}{{(u_I^2u_O^2 + {C_1})(\delta _I^2\delta _O^2 + {C_2})}},
%\end{align}
%\begin{align}\label{eq:chmd}
%  PSNR = 20lo{g_{10}}\left( {\frac{{{2^b} - 1}}{{\sqrt {MSE} }}} \right),
%\end{align}
where $\mathbf{I} = \left( {\mathbf{I}\left( {i,j} \right)} \right)$ is the input geometric location map, $\mathbf{O} = \left( {\mathbf{O}\left( {i,j} \right)} \right)$ is the corresponding CGM, ${{u_\mathbf{I}}}$ and ${{u_\mathbf{O}}}$ are the means, ${\delta _\mathbf{I}^2}$ and ${\delta _\mathbf{O}^2}$ are the variances of $\mathbf{I}$ and $\mathbf{O}$, respectively. ${{\delta _{\mathbf{IO}}}}$ is covariance of $\mathbf{I}$ and $\mathbf{O}$, $b$ is the number of bits stored for each pixel. $C_1$ and $C_2$ represent constants, avoiding 0 in the denominator \cite{4598839}.

Note that our work is based on a widely adopted dataset \cite{9354041}. Through analysis of this dataset, an important observation can be found in \figref{fig:lp}: In our task, we aim to obtain the channel gain for each spatial grid in the geometric location map, which is the difference between the geometric location map and the CGM. \figref{fig:lp} (a) shows a pair of original geometric location map and CGM, and \figrefs{fig:lp} (b)-(d) are their corresponding high-frequency components $\text{R} = [{\mathbf{r}_0},{\mathbf{r}_1}]$ and low-frequency component ${\mathbf{I}_2}$. Specifically, the MSE between the high-frequency components in \figrefs{fig:lp} (b)-(c) is smaller (about 7/15 and 2/5) than that between the low-frequency components in \figref{fig:lp} (d). In other words, the difference between the geometric location map and the CGM becomes more pronounced at low-frequency components than at high-frequency components. Similar observations can still be drawn from the variations in SSIM, PSNR and corresponding histograms, which means that the CGM reconstruction at low-frequency components is more significant. Consequently, we can reconstruct the differences on the low-frequency component with a scaled-down resolution, reducing largely the computational complexity. Note that during the high-resolution CGM reconstruction, it is essential to reasonably design the networks, based on the importance of different frequency components. %such as allocating more attention and computational resources to the low-frequency components with more differences to learn.

%\begin{figure*}[!t]
%\centering
%\includegraphics[scale=0.35]{figure/newlapmse.eps}
%%\includegraphics[width=0.3\textwidth]{figure/newlapmse.eps}
%\captionsetup{font=footnotesize}
%\caption{The LP decomposition obtains the difference between the geometric location map and the CGM for different frequency components. As shown by the MSE, SSIM, PSNR, and the Histogram, the differences between the geometric location map and the CGM are dominated by the low-frequency components.}
%\label{EE_ps}
%\end{figure*}

%?????????????????????????????????????????????????????%?????????????????????????????????????????????????????%?????????????????????????

\subsection{Laplacian Pyramid-Based CGM Network}
%?????????????????????????????????????????????????
%The inherent properties of LP, including the separation of textures and visual attributes, and the capability of a reversible reconstruction, can benefit our CGM reconstruction task. For traditional I2I methods, the specific attributes (differences) are represented in the latent space powered by a deep encoding-decoding network. In contrast, for the CGM reconstruction task, we observe that the difference can be extracted using fixed kernels in an efficient way. As shown in \figref{fig:lp}, for example, through the decomposition of the LP, the difference of the geometric location map-to-CGM reconstruction task is mainly exhibited in the low-frequency components. 

Based on the above findings, we are inspired to use the LP to decompose the original geometric location map, then design parallel sub-networks based on the frequency components with different importance. The CGM can be reconstructed by performing the corresponding inverse operation on the different frequency components. The proposed scheme demonstrates advantages over traditional I2I techniques in the following aspects. i) Low time and storage overhead: The decomposition of high- and low-frequency components in an LP is accomplished using a fixed kernel without learning from the image data. ii) Disentanglement and reconstruction effectiveness: The decomposition of components with different frequencies by an LP is a straightforward yet highly effective method for disentangling and reconstructing a targeted image, enhancing the overall reconstruction quality \cite{1095851}. On the contrary, traditional auto-encoders may face a trade-off between the number of model parameters and the effectiveness of disentanglement and reconstruction.
\begin{figure*}[!t]
	\centering
	\includegraphics[scale=0.32]{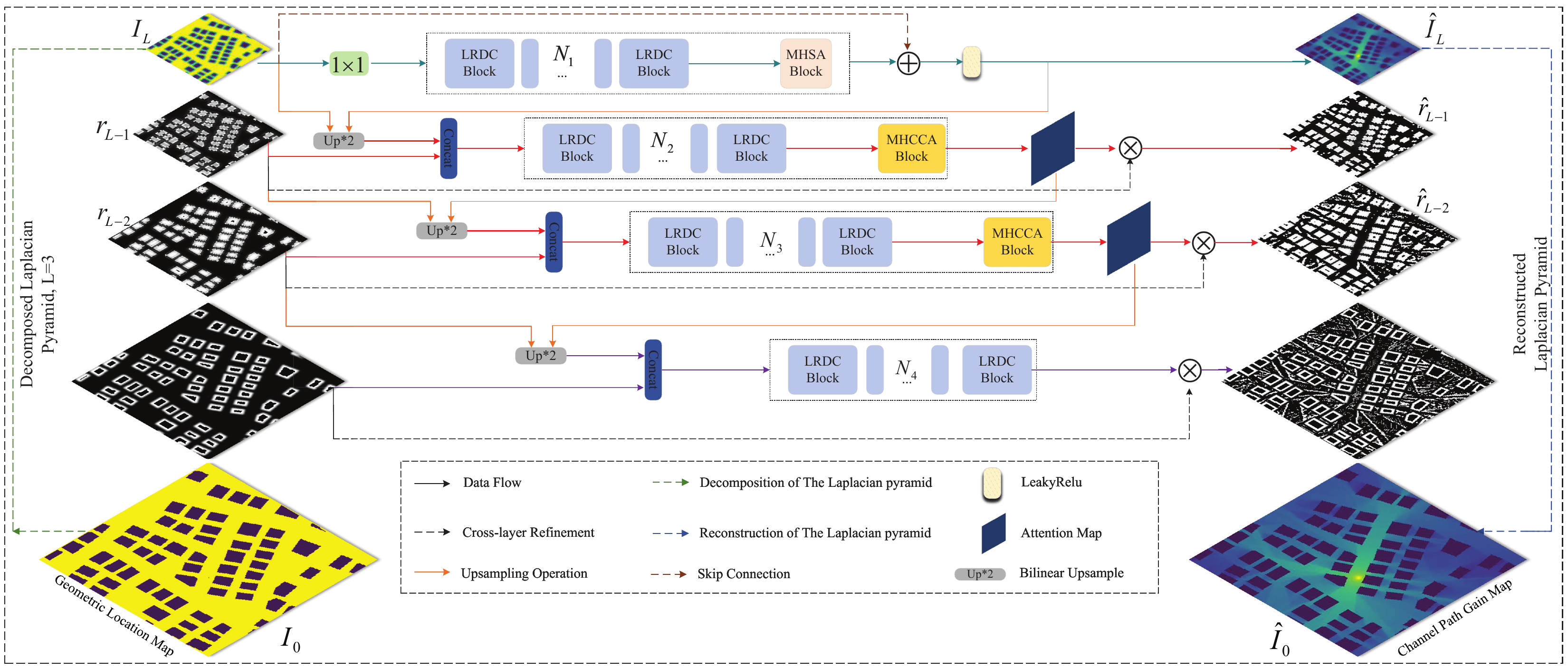}
	\captionsetup{font=footnotesize}
	\caption{Diagram of the framework of the proposed LPCGMN. When provided with a high-resolution morphological map ${\mathbf{I}_0} \in {\mathbb{R}^{h \times w \times c}}$, we initially decompose it using an $L$-level LP. \textcolor[RGB]{0,127,127}{Green arrow}: For the low-frequency component ${\mathbf{I}_L} \in {\mathbb{R}^{{\textstyle{h \over {{2^L}}}} \times {\textstyle{w \over {{2^L}}}} \times {\textstyle{c}}}}$, we reconstruct it into ${{\hat{\mathbf{I}}}_{L}}\in {{\mathbb{R}}^{\frac{h}{{{2}^{L}}}\times \frac{w}{{{2}^{L}}}\times c}}$ through the proposed lightweight network. \textcolor[RGB]{255,0,0}{Red arrow}: To adaptively refine the high-frequency component ${\mathbf{r}_{L - 1}} \in {\mathbb{R}^{\frac{h}{{{2^{L - 1}}}} \times \frac{w}{{{2^{L - 1}}}} \times c}}$, we learn an attention map  $\tilde{\boldsymbol{\psi}}_{Multihead}^{L-1}\in {{\mathbb{R}}^{\frac{h}{{{2}^{L-1}}}\times \frac{w}{{{2}^{L-1}}}\times 1}}$ based on both high- and low-frequency components. \textcolor[RGB]{112,48,160}{Purple arrows}: For the remaining components with higher resolutions, we progressively upsample the acquired attention map and refine it with the proposed LRDC blocks.}
	\label{fig:lpcgmn}
\end{figure*} 

Specifically, as shown in \figref{fig:lpcgmn}, given a geometric location map ${\mathbf{I}_0} \in {\mathbb{R}^{h \times w \times c}}$, where $h$, $w$, and $c$ are the height, width, and the number of color channels for the image, respectively. We first decompose it through the LP with $L$ levels, obtaining a set of high-frequency components defined by $\text{R} = [{\mathbf{r}_0},{\mathbf{r}_1},{\mathbf{r}_2},...,{\mathbf{r}_{L - 1}}]$ and a low-frequency geometric location submap ${\mathbf{I}_L} \in {\mathbb{R}^{\tfrac{h}{{{2^L}}} \times \tfrac{w}{{{2^L}}} \times c}}$. For the low-frequency geometric location components with more global information \cite{1095851}, we develop a deeper network, and integrate the attention mechanism to incorporate the global environment information into the features. Simultaneously, we progressively and adaptively refine the high-frequency components conditioned on the lower-frequency one.

（%the spatial relationships of the wireless communication environment can be extracted and represented in low-resolution geometric location components by designing deeper convolution and self-attention layers, and high-resolution components can be refined and globally adjusted by using low-resolution feature maps with global structure information. To avoid the problem of excessive convolution computation and time-consuming caused by directly processing a high-resolution geometric location map, it is necessary to reasonably allocate network resources.）
%根据以上的发现，我们利用拉普拉斯金字塔对原始几何位置地图进行分解，从而并行地对分解到的不同重要程度的频率成分分别设计子网络，最后CGM可以通过一系列镜像操作重建。具体地，对于具有更多全局信息的低频几何位置成分，我们设计了更深的卷积网络，并采用了注意力机制来将全局信息编码到局部特征中，完成局部-全局的交互，同时对高频几何位置成分自适应精修。In addition, we progressively refine the higher-resolution component conditioned on the lower-resolution one 
%According to the above findings, we use the Laplacian pyramid to decompose the original geometric location map, and then design sub-networks for the frequency components with different importance in parallel. Finally, the CGM can be reconstructed by a series of mirror operations.
%This subsection mainly studies and designs a lightweight feature extraction network, which can represent effective features and spatial relationships in different frequency components of the geometric location map.
%\subsubsection{Framework Overview of LPCGMN}
%Our main objective is to develop an efficient reconstruction model that can handle high-resolution geometric location maps for the CGM construction. 
Therefore, we propose an end-to-end learning framework called the Laplacian Pyramid-based CGM Reconstructed Network (LPCGMN), to reduce the computation complexity while maintaining competitive performance. The proposed LPCGMN is depicted in \figref{fig:lpcgmn}. The overall LPCGMN framework is mainly composed of two parts. \textit{i) Low-frequency} geometric location components: specially designed deeper convolutional network is used for feature (the spatial relationships of the wireless communication environment) representation and differences reconstruction. \textit{ii) High-frequency} geometric location components: tailored lightweight network for adaptive refinement and global/regional detail adjustment.

\textit{i) Reconstruction on Low-Frequency Component:} In general I2I tasks, reconstruction is often performed in a low-dimensional space using a series of cascaded residual blocks. Instead, we construct the CGM by leveraging the inherent properties of the LP. In short, the input is decomposed into submaps of different frequency bands, and their corresponding specific attributes are used to complete the feature extraction and reconstruction, which is beneficial to reduce the huge computational overhead in the traditional I2I task. As shown in \figref{fig:lpcgmn}, the low-frequency map ${\mathbf{I}_L}$ is first fed into a $1 \times 1$ convolution to complete the expansion of the dimension in the depth direction. Then the stacking of the proposed Lightweight Residual Dilated Convolutions (LRDC) module \textit{is used to deepen the feature extraction network and obtain the fusion of multi-scale features.} It is then fed into the proposed Multi-Head Self-Attention (MHSA) module\textit{ to encode random global structure information into local features in the spatial dimension, and learn more rich hierarchical feature representations.} Subsequently, we reduce the number of color channels in the feature maps to $c$ to obtain the result ${{\hat{\mathbf{I}}}_{L}}$.
%where $c$ denotes the number of channels of the input image.
%As can be seen from literature \cite{1095851}, the low-frequency component of Laplacian pyramid output is a fuzzy result, and each pixel of it is obtained by averaging neighboring pixels via an octave Gaussian filter. It can be found that ${I_L}$ reflects the global attribute features of the input image, which has the ability to guide the global content adjustment.

\textit{ii) Refinement on High-Frequency Component:} Since the low-frequency component of the LP output is a fuzzy result \cite{1095851}, ${\mathbf{I}_L}$ reflects the global attribute features of the input, which can guide the global content adjustment. Inspired by this, we fuse ${\mathbf{I}_L}$ and ${{\hat{\mathbf{I}}}_{L}}$ to guide the refinement of the high-frequency component progressively $\text{R} = [{\mathbf{r}_0},{\mathbf{r}_1},{\mathbf{r}_2},...,{\mathbf{r}_{L - 1}}]$, to obtain components with different frequencies for the reconstruction of the CGM. To begin with, the low-frequency geometric location submap ${\mathbf{I}_L} \in {\mathbb{R}^{{\textstyle{h \over {{2^L}}}} \times {\textstyle{w \over {{2^L}}}} \times {\textstyle{c}}}}$ and the predicted channel gain submap ${{\hat{\mathbf{I}}}_{L}} \in {\mathbb{R}^{{\textstyle{h \over {{2^L}}}} \times {\textstyle{w \over {{2^L}}}} \times {\textstyle{c}}}}$ are bilinearly upsampled respectively, and the corresponding spatial resolution is adjusted to ${\textstyle{h \over {{2^{L - 1}}}}} \times {\textstyle{w \over {{2^{L - 1}}}}}$, which is the same as the resolution of the high-frequency component ${r_{L - 1}}$ in the $\left( L - 1\right) $-th layer. Then, ${\mathbf{I}_L}$, ${{\hat{\mathbf{I}}}_{L}}$ and ${\mathbf{r}_{L - 1}}$ are concatenated and defined as ${\mathbf{G}^{L - 1}} \in {\mathbb{R}^{{\textstyle{h \over {{2^{L - 1}}}}} \times {\textstyle{w \over {{2^{L - 1}}}}} \times {\textstyle{c}}}}$. It is fed into the proposed LRDC module, and the output is denoted by $\mathbf{G}_{out}^{L - 1} \in {\mathbb{R}^{{\textstyle{h \over {{2^{L - 1}}}}} \times {\textstyle{w \over {{2^{L - 1}}}}} \times {\textstyle{c}}}}$. Then, the extracted multi-scale feature maps are fed into the proposed Multi-Head Cross-Covariance Attention (MHCCA) module \textit{to perform the attention along the feature dimension with a lower computational complexity.} Finally, the attention map for adjusting high-frequency component ${\mathbf{r}_{L - 1}}$ is obtained. Consequently, all high-frequency components of the LP can be gradually refined following the above steps, and the prediction results $\hat{\mathbf{R}} = [{\hat{ \mathbf{r}}_0},{\hat {\mathbf{r}}_1},{\hat {\mathbf{r}}_2},...,{\hat {\mathbf{r}}_{L - 1}}]$ of different components are obtained. With the help of ${\hat {\mathbf{I}}_L}$ and $\hat {\mathbf{R}} = [{\hat {\mathbf{r}}_0},{\hat {\mathbf{r}}_1},{\hat {\mathbf{r}}_2},...,{\hat {\mathbf{r}}_{L - 1}}]$, the ${\hat {\mathbf{I}}_0}$ is reconstructed by the inverse operation of the LP.
%modules
%\begin{figure}[!t]
%\centering
%\includegraphics[scale=0.35]{figure/LRDCnew.eps}
%%\includegraphics[width=0.35\textwidth]{figure/LRDCnew.eps}
%\captionsetup{font=footnotesize}
%\caption{Structures of the proposed Lightweight Residual Dilated Convolutions (LRDC) module.}
%\label{fig:lr}
%\end{figure}
\subsection{Key Modules in LPCGMN}
In this subsection, we mainly introduce the proposed three modules in LPCGMN.

\textit{i) Lightweight Residual Dilated Convolutions (LRDC) Module:} The LRDC module mainly utilizes depthwise separable convolution and dilated convolution, which can obtain multi-scale features with fewer model parameters. Depthwise separable convolution has several advantages over ordinary convolution: On one hand, the cross-channel information interaction and spatial correlation between features are enhanced. On the other hand, substantially decreasing the number of training parameters makes the model lightweight. Specifically, given the input with size $\left( {h_{\rm{in}}},{w_{\rm{in}}},{c_{\rm{in}}}\right) $, the output with size $\left( {h_{\rm{out}}},{w_{\rm{out}}},{c_{\rm{out}}}\right) $, and the convolution kernel with size $\left( k \times k\right) $. In terms of model parameters (also storage complexity), the number of parameters produced by the standard convolutional layer is given by ${N_{\rm{standard}}} = {c_{\rm{out}}} \times k \times k \times {c_{\rm{in}}}$. However, the number of parameters produced by depthwise separable convolution is written as ${N_{\rm{separable}}} = \left( {k^2} + {c_{out}}\right) {c_{\rm{in}}}$, and the ratio of the two can be expressed as
\begin{align}\label{eq:chmd}
{\textstyle{{{N_{\rm{separable}}}} \over {{N_{\rm{standard}}}}}} = {\textstyle{1 \over {{c_{\rm{out}}}}}} + {\textstyle{1 \over {{k^2}}}}.
\end{align} 
In terms of floating-point operations of the model (also time complexity), the standard convolutional layer is ${\cal O}\left( {k^2}{c_{\rm{in}}}{w_{\rm{out}}}{h_{\rm{out}}}{c_{\rm{out}}}\right) $ and the depthwise separable convolution layer is ${\cal O}\left( {k^2}{h_{\rm{in}}}{w_{\rm{in}}}{c_{\rm{in}}}\right)  + {\cal O}\left( {c_{\rm{in}}}{h_{\rm{out}}}{w_{\rm{out}}}{c_{\rm{out}}}\right) $. The ratio of the two can be given by
\begin{align}\label{eq:chmd}
\frac{{{{\cal O}_{\rm{separable}}}}}{{{{\cal O}_{\rm{standard}}}}} = \frac{{{h_{\rm{in}}}{w_{\rm{in}}}}}{{{h_{\rm{out}}}{w_{\rm{out}}}{c_{\rm{out}}}}} + \frac{1}{{{k^2}}}.
\end{align} 
With higher resolution of the input map, the values of (14) and (15) are small when using larger convolution kernels, which is beneficial for processing high-resolution maps. In addition, under the premise of a consistent number of model parameters, depthwise separable convolutions can make the model go deeper and obtain more advanced semantic information. 

The essence of CGM construction is to leverage the intrinsic spatial location relationship of these physical components, such as buildings, vehicles, trees, etc, within the propagation environment as a priori knowledge. Nevertheless, conventional convolutional neural networks (CNN) use convolutional and pooling layers to expand the range of receptive fields received by the next layer and simultaneously decrease the size of the feature map to a certain extent, which loses the spatial location information inside the input features \cite{sabour2017dynamic}. 
%Besides, related research by Princeton University also shows that modern neural networks applied to I2I generally use large-scale and continuous pooling and sub-sampling layers to fuse multi-scale background information, which not only seriously reduces the image's resolution but also loses the internal structure of the data \cite{yu2015multi}. 
%\begin{figure}[!t]
%	\centering
%	\includegraphics[scale=0.35]{figure/LRDCnew.eps}
%	%\includegraphics[width=0.35\textwidth]{figure/LRDCnew.eps}
%	\captionsetup{font=footnotesize}
%	\caption{Structures of the proposed Lightweight Residual Dilated Convolutions (LRDC) module.}
%	\label{fig:lr}
%\end{figure}
\begin{figure*}[!t]
	\centering
	\subfloat[LRDC module]{\includegraphics[width=0.26\textwidth]{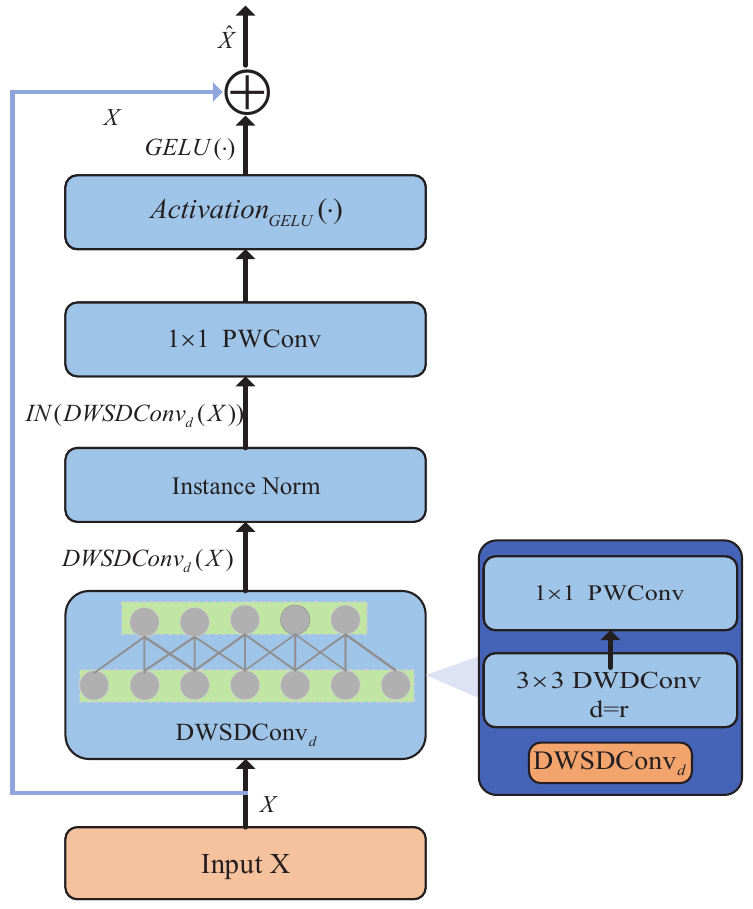}\label{fig:lr}}
	%\hspace{1in}
	\hspace{0.2in}
	%\hfill
	\subfloat[MHSA and MHCCA modules]{\includegraphics[width=0.62\textwidth]{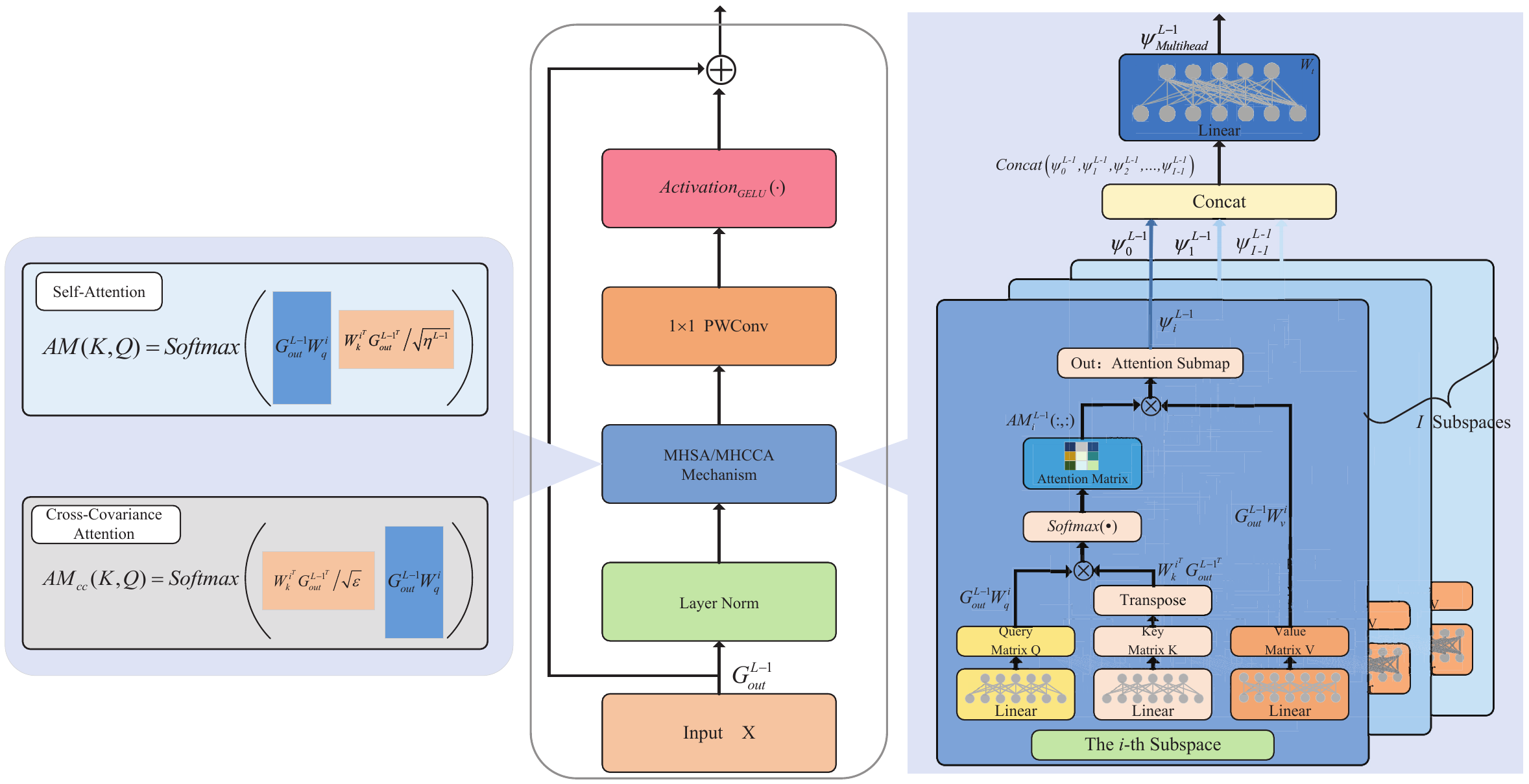}\label{fig:mhsa}}
	%newnewself.eps
	%MHnew.eps
	\captionsetup{font=footnotesize}
	\caption{Structures of the proposed Lightweight Residual Dilated Convolutions (LRDC), Multi-Head Self-Attention (MHSA) and Multi-Head Cross-Covariance Attention (MHCCA) modules.}
	\label{fig:module}
\end{figure*}
%To the best of our knowledge, conventional CNN usually uses convolutional and pooling layers to expand the range of receptive fields received by the next layer and simultaneously reduce the size of the feature map to a certain extent. Finally, using the upsampling operation to recover the size of the input image, but the loss of clarity inevitably occurs in the process of downsampling and upsampling of the feature map. 
Therefore, based on the depthwise separable convolution, we use successive dilated convolutions with different dilation rates for each layer instead of pooling operations. Without reducing the spatial resolution, the receptive field range also increases. In addition, it can capture multi-scale context information by controlling the different dilation rates of each layer. Note that no additional parameters are introduced in the whole dilated process since ``0'' is used to fill the convolution kernel in this paper. Specifically, let $d$ be defined as the dilated rate, the actual size of the dilated convolution kernel is $k' = k + \left( k - 1\right) \left( d - 1\right) $, and the receptive field of the $\left( i + 1\right) $-th layer is denoted as $R{F_{i + 1}}$, which is given by 
\begin{align}\label{eq:chmd}
{S_i} = \sum\nolimits_{i = 1}^I {\rm{stride}_{\mathit{i}}},
\end{align}
\begin{align}\label{eq:chmd}
R{F_{i + 1}} = R{F_i} + \left( k' - 1\right)  \times {S_i},
\end{align}
where ${S_i}$ is the product of all strides in the previous $(I-1)$ layers and $R{F_i}$ is defined as the receptive field at the $i$-th layer. When computing the receptive field at the $\left( i + 1\right) $-th layer, $R{F_1}$ is initialized to 1 by default. Let $x[n]$ be defined as the input, and the output is denoted as
\begin{align}\label{eq:chmd}
y[n] = \sum\nolimits_{k = 1}^K {x[n + k \cdot d]} w[k],
\end{align}
where $w[k]$ is a filter with length $K$. Considering an input feature $\mathbf{X}$ with dimension $H \times W \times C$, The proposed LRDC module outputs $\hat {\mathbf{X}}$ is given by
\begin{align}\label{eq:chmd}
\hat {\mathbf{X}} = \mathbf{X} + \rm{GELU}\left( {\rm{PWConv}\left( {\rm{IN}\left( {{\rm{DWSDConv}_\textit{d}}\left( \mathbf{X} \right)} \right)} \right)} \right),
\end{align}
where $\rm{IN}\left(  \cdot \right) $ is an instance normalization layer, $\rm{GELU}\left(  \cdot \right) $ is the GELU activation, ${\rm{PWConv}}\left(  \cdot \right) $ is a pointwise convolution, ${\rm{DWSDConv}_\textit{d}}\left(  \cdot \right) $ is a $3 \times 3$ depthwise separable dilated convolution with dilated rate $d$. The procedure of the LRDC module is summarized in \subfigref{fig:module}{fig:lr}.

\textit{ii) The Multi-Head Self-Attention (MHSA) Module:} Essentially, the self-attention mechanism functions as a neural network module capable of implementing a parallel input feature weighting function, which is advantageous to encode global structure information into the features \cite{vaswani2017attention}. To implement the self-attention mechanism, a normalized attention matrix is introduced to represent varying degrees of attention to the input. More significant input components are assigned higher weights. The final output is computed by weighting the input based on the attention weights specified in the attention matrix. With this mechanism, the input features can generate a novel feature representation, carrying the global structure information. To make this paper coherent, let's first briefly review how self-attention mechanism works. Given the input $\mathbf{Z} = {[{\mathbf{z}_1},...,{\mathbf{z}_m}]^T} \in {\mathbb{R}^{{d_m} \times {d_n}}}$, ${{d_m}}$ denotes the number of image patches and ${{d_n}}$ is the feature dimension of each image patch, three different linear transformation are applied to ${\mathbf{z}_j}$ \cite{vaswani2017attention}:
\begin{subequations}\label{eq:spdb}
\begin{align}
	{\mathbf{k}_j} &= {\mathbf{z}_j}{\mathbf{W}^k},\quad j = 1,...,{d_n},\label{eq:dbso}\\
	{\mathbf{q}_j} &= {\mathbf{z}_j}{\mathbf{W}^q},\quad j = 1,...,{d_n},\label{eq:dbsc}\\
	{\mathbf{v}_j} &= {\mathbf{z}_j}{\mathbf{W}^v},\quad j = 1,...,{d_n},\label{eq:dbsc}
\end{align}
\end{subequations}
where ${\mathbf{k}_j} \in {\mathbb{R}^{1 \times {d_k}}}$, ${\mathbf{q}_j} \in {\mathbb{R}^{1 \times {d_q}}}$ and ${\mathbf{v}_j} \in {\mathbb{R}^{1 \times {d_n}}}$ are the key, query and value vector, respectively. ${\mathbf{W}^k} \in {\mathbb{R}^{{d_n} \times {d_k}}}$, ${\mathbf{W}^q} \in {\mathbb{R}^{{d_n} \times {d_q}}}$ and ${\mathbf{W}^v} \in {\mathbb{R}^{{d_n}\times {d_n}}}$ represent the respective trainable transformation matrices, with ${d_k} = {d_q}$. In detail, the weight allocation function is determined by ${\mathbf{k}_j}$ and ${\mathbf{q}_{j'}}$. The correlation between ${\mathbf{k}_j}$ and ${\mathbf{q}_{j'}}$ represents the correlation between the $j$-th geometric location map patch and the ${j'}$-th CGM patch. A higher correlation ${\mathbf{q}_{j'}}\mathbf{k}_j^T$ implies that the features of the $j$-th input patch ${\mathbf{z}_j}$ hold greater importance for the ${j'}$-th output patch. Generally, this correlation can be adaptively adjusted based on the input $\mathbf{Z}$ and the matrices ${\mathbf{W}^k}$ and ${\mathbf{W}^q}$. For clarity, the matrix forms of (20a)-(20c) are presented as follows:
\begin{subequations}\label{eq:spdb}
\begin{align}
	{\mathbf{K}} &= {\mathbf{Z}}{\mathbf{W}^k},\label{eq:dbso}\\
	{\mathbf{Q}} &= {\mathbf{Z}}{\mathbf{W}^q},\label{eq:dbsc}\\
	{\mathbf{V}} &= {\mathbf{Z}}{\mathbf{W}^v},\label{eq:dbsc}
\end{align}
\end{subequations}
where $\mathbf{K} = {[{\mathbf{k}_1},...,{\mathbf{k}_m}]^T} \in {\mathbb{R}^{{d_m} \times {d_k}}}$, $\mathbf{Q} = {[{\mathbf{q}_1},...,{\mathbf{q}_m}]^T} \in {\mathbb{R}^{{d_m} \times {d_q}}}$ and $\mathbf{V} = {[{\mathbf{v}_1},...,{\mathbf{v}_m}]^T} \in {\mathbb{R}^{{d_m} \times {d_n}}}$.
 
Using $\mathbf{K}$ and $\mathbf{Q}$, we can obtain the attention matrix $\mathbf{AM}  \in {\mathbb{R}^{{d_m} \times {d_m}}}$, which is denoted as 
\begin{align}\label{eq:chmd}
\mathbf{AM} = \rm{Softmax} \left( {\frac{{\mathbf{Q}{\mathbf{K}^\textit{T}}}}{{\mathit{\sqrt \eta} }}} \right),
%\mathbf{AM} = \rm Softmax \left( {\frac{{\mathbf{Q}{\mathbf{K}^T}}}{{\sqrt {\textit{d}} }}} \right),
\end{align}
where $\rm{Softmax} \left( \mathbf{Z}\right)  = \frac{{\exp \left( \mathit{{z_j}}\right) }}{{\sum {\exp \left( \mathit{{z_j}}\right) } }}$, $\sqrt \eta$ is a scaling factor. Each column of the attention matrix is a vector of attention scores, i.e., each score is a probability, where all scores are non-negative and sum up to 1. Note that when the key vector $\mathbf{K}[j,:]$ and the query $\mathbf{Q}[j',:]$ have a better match, the corresponding attention score $\mathbf{AM}[j',j]$ is higher. Thus, the output of the attention mechanism corresponding to the $r$-th component can be represented by the weighted sum of all inputs, denoted as \cite{9832933}
\begin{align}\label{eq:chmd}
{\mathbf{o}_r} = \sum\limits_j {\mathbf{AM} [r,j]{\mathbf{v}_j}}  = \mathbf{AM} [r,:] \cdot \mathbf{V},
\end{align}
where ${\mathbf{o}_r} \in {\mathbb{R}^{1 \times {d_n}}}$ represents the $r$-th output, which is computed by adaptively focusing on the inputs based on the attention score $\mathbf{AM}[r,j]$. When the attention score $\mathbf{AM}[r,j]$ is higher, the associated value vector ${\mathbf{v}_j}$ will have a more significant impact on the $r$-th output patch. Finally, the overall representation of the attention map is given by 
\begin{align}\label{eq:chmd}
\mathbf{O} &=\mathbf{AM}[:,:]\cdot \mathbf{V}\nonumber \\
&=\rm{Softmax}\mathit{\left( {\frac{{\mathbf{Z}{\mathbf{W}^q}{\mathbf{W}^{{k^T}}}{\mathbf{Z}^T}}}{{\sqrt \eta }}} \right)\mathbf{Z}{\mathbf{W}^v}},
\end{align}
where $\mathbf{O} = {[{\mathbf{o}_{1,...,}}{\mathbf{o}_m}]^T} \in {\mathbb{R}^{{d_m} \times {d_n}}}$.
%\begin{figure*}[!t]
%\centering
%%\vspace{-0.1cm}%%减小图片上间隔
%\includegraphics[scale=0.3]{figure/MHnew.eps}
%%\includegraphics[width=0.8\textwidth]{figure/MHnew.eps}
%\captionsetup{font=footnotesize}
%%\vspace{-0.1cm}%%减小图片上间隔
%\caption{Structures of the proposed Multi-Head Self-Attention (MHSA) Module and  Multi-Head Cross-Covariance Attention (MHCCA) Module.}
%%\vspace{0.01cm}%%减小图片上间隔
%\label{fig:mhsa}
%\end{figure*}

Based on the above, we present the proposed MHSA module for CGM construction. Specifically, given the output of the LRDC module at the $\left( L-1\right) $-th layer as $\mathbf{G}_{\rm{out}}^{L - 1} \in {\mathbb{R}^{\hat H \times \hat W \times \hat C}}$, which is first fed into the normalization layer and then linearly transformed to the queries $\mathbf{Q} = \mathbf{G}_{\rm{out}}^{L - 1}{\mathbf{W}^q}$, keys $\mathbf{K} = \mathbf{G}_{\rm{out}}^{L - 1}{\mathbf{W}^k}$, and values $\mathbf{V} = \mathbf{G}_{\rm{out}}^{L - 1}{\mathbf{W}^v}$.
%where ${W_{dw}}$ and ${W_{pw}}$ are weight matrices. ${W_{pw}}$ is the $ 1\times 1$ point-wise convolution used to aggregate pixel-wise cross-channel context. ${W_{dw}}$ is the $ 3\times 3$ depth-wise convolution used to encode channel-wise spatial context. 
Then, the single-head self-attention map is obtained by
\begin{align}\label{eq:chmd}
\boldsymbol{\psi}  = \rm{Softmax}\mathit{\left( {\frac{{\mathbf{G}_{\rm{out}}^{L - 1}{\mathbf{W}^q}{\mathbf{W}^{{k^T}}}\mathbf{G}_{\rm{out}}^{L - {1^T}}}}{{\sqrt \eta }}} \right)\mathbf{G}_{\rm{out}}^{L - 1}{\mathbf{W}^v}}.
\end{align}
According to (24), it can be extended to a multi-head self-attention learning mechanism. Firstly, the learned self-attention submap in the $i$-th subspace is calculated as follows:
\begin{subequations}\label{eq:spdb}
\begin{align}
	&\tilde {{\mathbf{G}}}_{\rm{out},\mathit{i}}^{L - 1} = \mathbf{G}_{\rm{out}}^{L - 1}{\mathbf{W}^i},\label{eq:dbso}\\
	\boldsymbol{\psi} _i^{L - 1} &= \rm{Softmax}\mathit{\left( {\frac{{\tilde {\mathbf{{G}}}_{\rm{out},\mathit{i}}^{L - 1}\tilde {\mathbf{{G}}}_{\rm{out},\mathit{i}}^{L - {1^T}}}}{{\sqrt {{\eta^{l - 1}}} }}} \right)},\label{eq:dbsc}
\end{align}
\end{subequations}
where $\tilde {\mathbf{{G}}}_{\rm{out},\mathit{i}}^{L - 1}$ is the mapping of $\mathbf{G}_{\rm{out}}^{L - 1}$ in the $i$-th learned subspace. ${\mathbf{W}^i}$ is a trainable transformation matrix, which is denoted as
\begin{align}\label{eq:chmd}
{\mathbf{W}^i} = \left\{ \begin{gathered}
	{\mathbf{W}_i}^q,\text{if}{\text{ }}{\mathbf{W}^i}{\text{ }}{\text{is the linear transformation of }}\mathbf{Q} \hfill \\
	{\mathbf{W}_i}^k,\text{else} \hfill \\ 
\end{gathered}  \right.,
\end{align}
and it can map $\mathbf{G}_{\rm{out}}^{L - 1}$ into different self-attention learning subspaces to form an I-head self-attention map. 
%To further enhance the information interaction between different spatial features and improve the model's ability to learn global information, as well as increase operational efficiency, we use full self-attention. Full self-attention implies that Eq. (25) satisfies the following conditions: $Q = \tilde G_{out,i}^{L - 1}$ and ${K^T} = \tilde G_{out,i}^{L - {1^T}}$.

Then, the self-attention submaps of the I transformed subspaces are concatenated as follows:
\begin{subequations}\label{eq:spdb}
\begin{align}
	\boldsymbol{\psi} _{\rm{Multihead}}^{L - 1} = \rm{Concat}\mathit{\left( {\boldsymbol{\psi} _0^{L - 1},\boldsymbol{\psi} _1^{L - 1},\boldsymbol{\psi} _2^{L - 1},...,\boldsymbol{\psi} _{I-1}^{L - 1}} \right){\mathbf{W}_t}},\label{eq:dbso}\\
	\tilde{\boldsymbol{{\psi }}}_{\rm{Multihead}}^{L-1}=\rm{GELU}\mathit{\left( \rm{PWConv}\left( \tilde{\boldsymbol{{\psi}}}_{Multihead}^\mathit{{L-1}} \right) \right)+\mathbf{G}_{\rm{out}}^{L-1}},\label{eq:dbsc}
\end{align}
\end{subequations}
where ${\mathbf{W}_t}$ is a trainable matrix, $\tilde {\boldsymbol{{\psi}}} _{\rm{Multihead}}^{L - 1} \in {\mathbb{R}^{^{{\textstyle{h \over {{2^{L - 1}}}}} \times {\textstyle{w \over {{2^{L - 1}}}}} \times 1}}}$, $i \in [1,I]$, $I$ is the number of parallel self-attention learning subspaces or heads. The obtained self-attention map $\tilde {\boldsymbol{{\psi}}} _{\rm{Multihead}}^{L - 1}$ is used for pixel-level refinement of the high-frequency submap ${\mathbf{r}_{L - 1}}$:
\begin{align}\label{eq:chmd}
{\hat {\mathbf{{r}}}_{L - 1}} = {\mathbf{r}_{L - 1}} \otimes \tilde {\boldsymbol{{\psi}}} _{\rm{Multihead}}^{L - 1},
\end{align}
where $ \otimes $ is the multiplication at the pixel level. Through the above steps, we complete the pixel-level refinement of the high-frequency component ${\mathbf{r}_{L - 1}} \in {\mathbb{R}^{^{{\textstyle{h \over {{2^{L - 1}}}}} \times {\textstyle{w \over {{2^{L - 1}}}}} \times 1}}}$ at the $\left( L-1\right)$-th layer of the LP.

For the refinement of the remaining frequency components $\mathbf{R} = [{\mathbf{r}_0},{\mathbf{r}_1},{\mathbf{r}_2},...,{\mathbf{r}_{L - 2}}]$, extend $\tilde {\boldsymbol{{\psi}}} _{\rm{Multihead}}^{L - 1}$ to the corresponding self-attention maps $[\tilde {\boldsymbol{{\psi}}} _{\rm{Multihead}}^{L - 2},...,\tilde {\boldsymbol{{\psi}}} _{\rm{Multihead}}^1,\tilde {\boldsymbol{{\psi}}} _{\rm{Multihead}}^0]$ to the rest frequency components through the proposed upsampling operation, LRDC and MHCCA module to obtain the refined results of the remaining frequency components $[{\hat {\mathbf{{r}}}_{L - 1}},{\hat {\mathbf{{r}}}_{L - 2}},...,{\hat {\mathbf{{r}}}_1},{\hat {\mathbf{{r}_0}}}]$, where $\tilde {\boldsymbol{{\psi}}} _{\rm{Multihead}}^{L - 2} \in {\mathbb{R}^{{\textstyle{h \over {{2^{L - 2}}}}} \times {\textstyle{w \over {{2^{L - 2}}}}} \times 1}}$ and $\tilde {\boldsymbol{{\psi}}} _{\rm{Multihead}}^0 \in {\mathbb{R}^{h \times w \times 1}}$. Then, with the help of the reversibility of LP, the CGM is obtained using the reconstructed ${{\hat {\mathbf{{I}}}}_L}$ and the refined $[{\hat {\boldsymbol{{r}}}_{L - 1}},{\hat {\boldsymbol{{r}}}_{L - 2}},...,{\hat {\boldsymbol{{r}}}_1},{\hat {\mathbf{{r}}}_0}]$. The procedure of the MHSA module is summarized in \subfigref{fig:module}{fig:mhsa}.

%And then completing the reconstruction of the channel path gain map according to the characteristics of the LP.

\textit{iii) The Multi-Head Cross-Covariance Attention (MHCCA) Module:} For the high-frequency components $\text{R} = [{\mathbf{r}_0},{\mathbf{r}_1},{\mathbf{r}_2},...,{\mathbf{r}_{L - 2}}]$, the MHSA module is no longer applicable, since it is accompanied by quadratic complexity of time and storage, which hinders the application of high-resolution components. The proposed MHCCA module utilizes the cross-covariance attention mechanism \cite{ali2021xcit} to compute the attention score along the feature dimension rather than across the spatial (or token) dimension. Compared to the MHSA, it decreases the storage complexity from ${\cal O}\left( {Id_m^2 + {d_m}{d_n}} \right)$ to ${\cal O}\left( {{{d_n^2} \mathord{\left/
			{\vphantom {{d_n^2} I}} \right.
			\kern-\nulldelimiterspace} I} + {d_m}{d_n}} \right)$, and reduces the time complexity from ${\cal O}\left(  d_m^2{d_n}\right)  $ to ${\cal O}\left(  {{d_m{d_n^2}} \mathord{\left/
	{\vphantom {{d_m^2{d_n}} I}} \right.
	\kern-\nulldelimiterspace} I}\right)  $. Due to the quadratic complexity of time, it is problematic to extend self-attention to higher-resolution images. The MHCCA overcomes this drawback as its time complexity increases linearly with the number of image patches, and so does its storage complexity. The attention operation in the MHCCA module is a transposed version of self-attention based on the cross-covariance matrix between keys and queries, which is denoted as
\begin{subequations}\label{eq:spdb}
\begin{align}
	{{\mathbf{AM}}_{\rm{cc}}}&\left( \mathbf{Q},\mathbf{K}\right) =\rm{Softmax}\mathit{\left( \frac{{{\mathbf{K}}^{T}}\mathbf{Q}}{\varepsilon } \right)},\label{eq:dbso}\\
	&{\boldsymbol{\psi} _{\rm{cc}}} = \mathbf{V} \cdot {{\mathbf{AM}}_{\rm{cc}}}\left( \mathbf{Q},\mathbf{K}\right) ,\label{eq:dbsc}
\end{align}
\end{subequations}
%\begin{align}\label{eq:chmd}
%	\mathord{\buildrel{\lower3pt\hbox{$\scriptscriptstyle\smile$}} 
%		\over X} = CC\_Atteneion(Q,K,V),
%\end{align}
%\begin{align}\label{eq:chmd}
%    CC\_Atteneion(Q,K,V) = V \cdot {\mathop{\rm softmax}\nolimits} ({{{Q^T} \cdot K} \mathord{\left/
	%    		{\vphantom {{{Q^T} \cdot K} \varepsilon }} \right.
	%    		\kern-\nulldelimiterspace} \varepsilon }),
%\end{align} 
where ${{\mathbf{AM}}_{\rm{cc}}}\left( \mathbf{Q},\mathbf{K}\right) $, ${\boldsymbol{\psi} _{\rm{cc}}}$ are the cross-covariance attention matrix and map, respectively. $\varepsilon $ is a scaling factor. Apart from this, the MHCCA module is identical to the previously mentioned MHSA module. The procedure of the MHCCA module is summarized in \subfigref{fig:module}{fig:mhsa}.
%Then, the MHCCA module outputs $\hat X$ is given by
%\begin{align}\label{eq:chmd}
%	\hat X = GE((DSDCon{c_r}(IN(\mathord{\buildrel{\lower3pt\hbox{$\scriptscriptstyle\smile$}} 
%		\over X}  + X)))) + X,
%\end{align} 
%where $GE( \cdot )$ is the GELU activation. 
\section{Numerical Experiment}\label{sec:sim}
In this section, we conduct a series of experiments to verify the performance of the proposed network architecture and modules. We first introduce the datasets and the experiment setup. Then, to explore the influence of different hyperparameter settings, we first conduct hyperparameter optimization experiments. Finally, we compare our scheme with the Unet and WNet-based CGM construction schemes regarding CGM reconstruction accuracy, time complexity, storage complexity, and robustness.
\subsection{Datasets}
\begin{figure}[!b]
\centering
\includegraphics[scale=0.65]{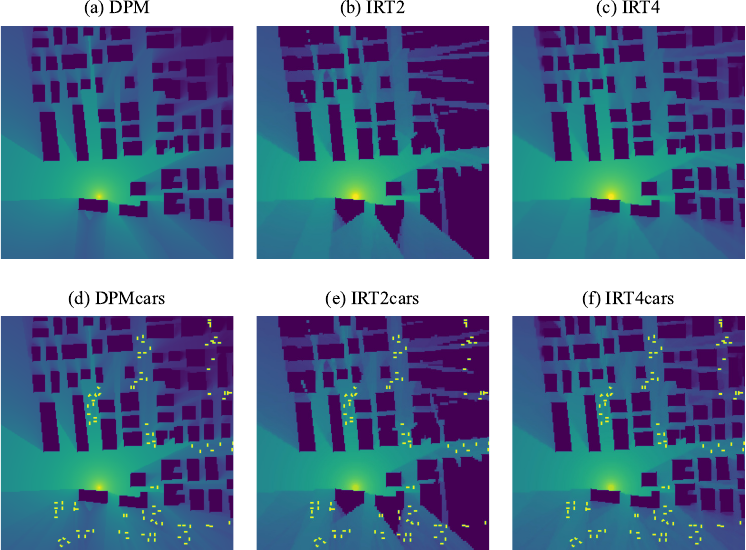}
\captionsetup{font=footnotesize}
\caption{Partial samples of DPM, DPMcars, IRT2, IRT2cars, IRT4, IRT4cars datasets.}
\label{fig:data}
\end{figure}

%为了进一步验证我们所提出的LPCGMN框架是否可以高效的构建信道增益地图，实验部分将采用RadioMapSeer数据集[1]来进行模型的训练和验证。The RadioMapSeer dataset主要包括了700张城市地图，每张地图包含了80个发射器的位置，及其相对应模拟的信道路径增益地图。其中，城市地图主要由OpenStreetMap所提供，涉及到的城市包括了London，Berlin，Glasgow等。按照数据集中信道路径增益地图模拟的方式，其大致分为了两类，The coarse simulations are generated using the Dominant Path Model (DPM) method [26] and Intelligent Ray Tracing (IRT) [53] based on 2 interactions of the rays with the geometry. Moreover, the elaborate simulations are generated using the IRT with 4 interactions (IRT4), for the first two transmitters of each city map.以上粗糙的模拟结果和精细的模拟结果均是由WinProp【】计算得到，关于系统的具体参数设置如表格1所示。所采用的数据集可细分如以下几个：
The RadioMapSeer dataset \cite{9354041} is used to train and verify our model. The RadioMapSeer dataset consists of 700 city maps, each containing the locations of 80 transmitters and their corresponding simulated channel gains. The city map is mainly provided by OpenStreetMap \cite{9354041}, and the cities involved include London, Berlin, Glasgow, etc. According to different generation algorithms, three corresponding datasets can be obtained. i) DPM and ii) IRT2 are coarse CGM datasets generated using the DPM and IRT2 algorithms, respectively. iii) IRT4 is a fine CGM dataset generated using the IRT4 algorithm. They are all provided by the software WinProp. To increase the uncertainty of the datasets, the influence of cars is added to the city map, and the above simulation algorithms and WinProp software are used to generate the corresponding datasets, namely DPMcars, IRT2cars, and IRT4cars. Examples of the six datasets mentioned above are shown in \figref{fig:data}. To ensure data quality and consistency, we conduct relevant data cleaning and preprocessing operations on the collected data, which mainly involves processing missing values, outliers, and duplicate records, removing noise, and normalizing images. Additionally, to ensure diversity in the training data and enhance the model's generalization, we employ data augmentation techniques, mainly rotating and flipping the images to generate a more diverse set of training samples. All datasets are saved as a dense sampling of the channel knowledge map in a 2D grid of $256 \times 256$ $\rm{m}^2$.
%为了增加数据集的不确定性[]，通过在城市地图中加入车辆的影响，进而借助以上三种模拟算法以及WinProp软件生成所对应的三个数据集，分别为DPMcars，IRT2cars，IRT4cars。关于以上所提到六个数据集的例子被展示在图3中。在信道路径增益地图中，每一个最小空域位置与放射器位置间的路径增益以dB的形式储存。

%经典三线表
\newcolumntype{L}{>{\hspace*{-\tabcolsep}}l}
\newcolumntype{R}{c<{\hspace*{-\tabcolsep}}}
\definecolor{lightblue}{rgb}{0.93,0.95,1.0}
\begin{table}[!t]
\captionsetup{font=footnotesize}
\caption{System and Model Parameters}\label{System Parameters}
\centering
\setlength{\tabcolsep}{13mm}%宽度
\ra{1.5}%高度
\scriptsize
%\rowcolors{1}{lightblue}{white}
\scalebox{0.8}{\begin{tabular}{LR}
		%\scriptsize
		\toprule
		Parameter &  Value\\
		\midrule
		%Earth radius $R_e$ && 6378 $(\text{km})$\\
		\rowcolor{lightblue}
		Minimum interval units & ${\Delta _x}=1$ m   \\
		Minimum interval units & ${\Delta _y}=1$ m    \\
		\rowcolor{lightblue}
		Number of rows of the grid & ${N_x}=256$\\
		Number of colums of the grid & ${N_y}=256$\\
		\rowcolor{lightblue}
		Number of transmitters & 80\\
		Carrier frequency & $f=5.9$ GHZ\\
		\rowcolor{lightblue}
		Bandwidth & $B=10$ MHZ\\
		Transmit power & 23 dBm\\
		\rowcolor{lightblue}
		Noise figure & 0 dB\\
		Noise power sepctral density & -174 dBm/HZ\\
		\rowcolor{lightblue}
		Optimizer & Adam\\
		The number of levels of the LP & $L = [1,2,3,4,5,6,7,8]$\\
		\rowcolor{lightblue}
		The dilation rate & $d = \{ 1,2,3\} $\\
		The number of the LRDC block & $\{ {N_1} = 4,{N_2} = 3,{N_3} = 3,{N_4} = 2\} $\\
		\rowcolor{lightblue}
		The number of heads of the attention  & $I = \{ 1,2,3,4,5,6,7\} $\\
		Number of training & 40000\\
		\rowcolor{lightblue}
		Number of testing & 8000\\
		Number of validating & 8000\\
		\bottomrule
	\end{tabular}
}
\end{table}

\subsection{Experiment Setup}
This paper has some common settings among the experiments involved: On the hardware level, the model is trained on Nvidia RTX-4090 GPU with 24 GB of memory. On the algorithm level, we employ Adam optimizer with learning rate ${10}^{ - 4}$. The proposed LPCGMN is trained in a supervised mode by minimizing the reconstruction loss ${\cal L}{}_{{\rm{recons}}}$, where ${\cal L}{}_{{\rm{recons}}}$ is the MSE between the reconstructed CGM and the desired CGM from the training set. Note that the 56,000 samples are randomly divided into the corresponding training, validation, and test sets in a ratio of 5:1:1. The training epochs are 50, and the batch size is set to 16. To ensure consistent training and evaluation criteria as in \cite{9354041}, we also select the model with the lowest ${\cal L}{}_{{\rm{recons}}}$ in the validation set from 50 epochs to avoid overfitting. Finally, the selected model is evaluated on a coarse simulation or IRT4 simulation of the test set. Note that RMSE evaluates the LPCGMN model performance on the gray levels and by NMSE. More detailed system and model parameters are summarized in Table I.

%经典三线表
%\begin{table}[!t]
%\captionsetup{font=footnotesize}
%\caption{Parameters Setting for the Proposed LPCGMN}\label{Parameters setting for the proposed LPCGMN}
%\centering
%\setlength{\tabcolsep}{12mm}%宽度
%\ra{1.3}%高度
%\scriptsize
%%\rowcolors{1}{lightblue}{white}
%\scalebox{0.8}{\begin{tabular}{LR}
%		%\scriptsize
%		\toprule
%		Parameters &  Value\\
%		\midrule
%		%Earth radius $R_e$ && 6378 $(\text{km})$\\
%		\rowcolor{lightblue}
%		%Batch size & 15   \\
%		Learning rate & $1{e^{ - 4}}$    \\
%		\rowcolor{lightblue}
%		Training epochs & 50\\
%		Loss function & MSE\\
%		\rowcolor{lightblue}
%		Optimizer & Adam\cite{amda}\\
%		The number of levels of the LP & $L = [3,4,5]$\\
%		\rowcolor{lightblue}
%		The dilation rate & $d = \{ 1,2,3\} $\\
%		The number of the LRDC block & $\{ {N_1} = 4,{N_2} = 3,{N_3} = 3,{N_4} = 2\} $\\
%		\rowcolor{lightblue}
%		the number of heads of the attention  & $I = \{ 1,2,3,4,5,6,7\} $\\
%		Number of training  & 40000\\
%		\rowcolor{lightblue}
%		Number of testing & 8000\\
%		Number of validating & 8000\\
%		\bottomrule
%	\end{tabular}
%}
%\end{table}

\subsection{Experiment Results}
%In this subsection, we first conduct hyperparameter optimization experiments to explore the influence of different hyperparameter settings, including the level of LP and the number of attention heads, on the reconstruction performance of CGM, then make a more comprehensive comparison between the model under the optimal hyperparameter setting and the current advanced methods.
\subsubsection{Experiment Results on LPCGMN with Different Levels of LP}
\begin{figure*}[!t]
\centering 
\includegraphics[scale=0.351]{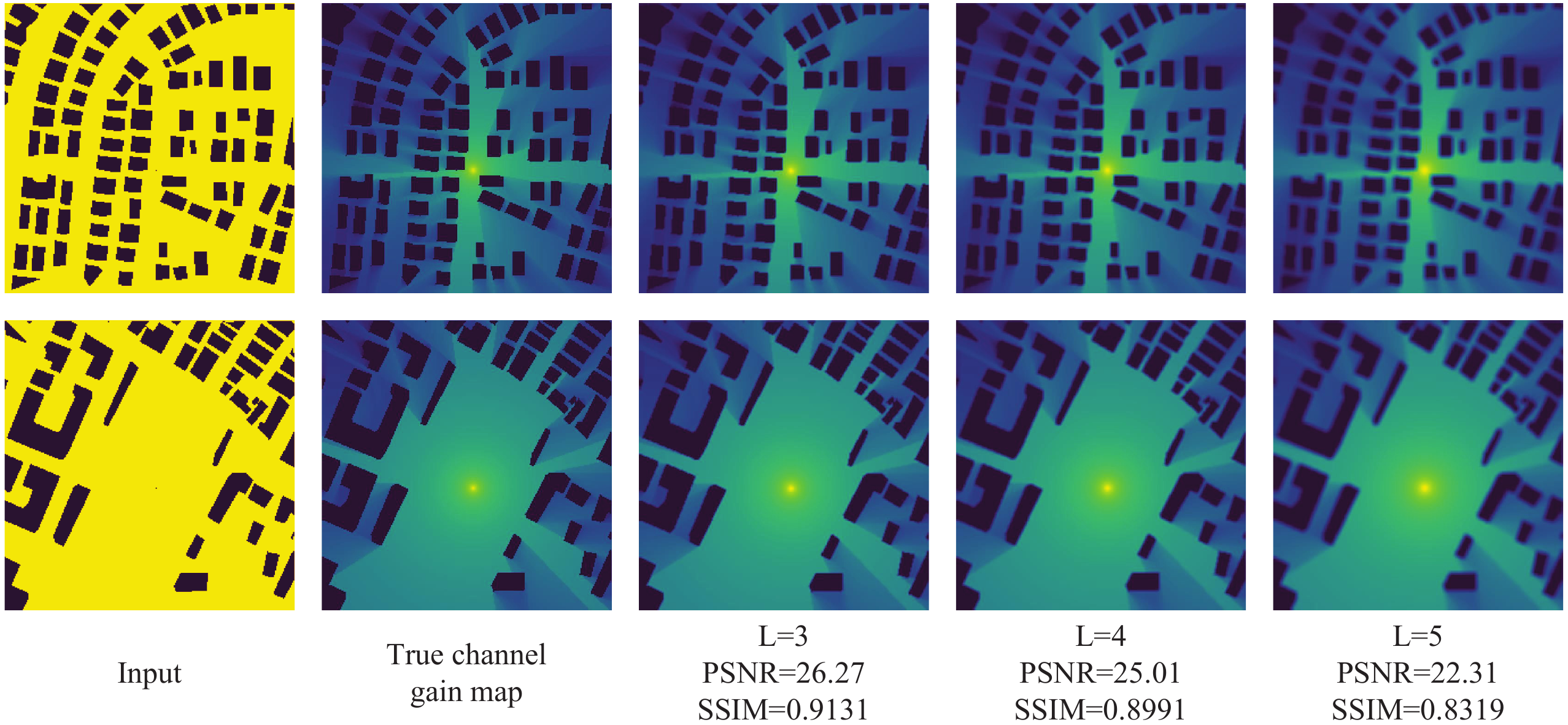}
\captionsetup{font=footnotesize}
\caption{Performance comparison of different numbers of LP layers of the proposed LPCGMN on DPM.}
\label{fig:psnr}
\end{figure*}

We first explore the correlation between the number of LP layers and the prediction performance of the proposed LPCGMN in which the number $I$ of attentional heads in the MHCCA and MHSA modules is set to 5. The three cases of $L =3$, $L =4$, and $L =5$ are validated during the LPCGMN training and testing, respectively. As shown in \figref{fig:psnr}, we visualize the reconstructed CGMs and their corresponding PSNR and SSIR evaluation metrics for the three parameter settings. Note that with the increase in the number of levels of the LP, the difference between the CGM reconstructed by the LPCGMN and the real one increases. When $L$ increases from 3 to 5, the corresponding PSNR decreases from 26.27 to 22.31, and SSIM decreases from 0.9131 to 0.8319. The number of LP levels impacts the trade-off between time consumption and reconstruction accuracy. From the point of view of time consumption, the RadioUnet \cite{9354041} takes about 16 m 41 s to train an epoch on the DPM (8 m 30 s for the first Unet and 8 m 11s for the second Unet). However, the corresponding time for the proposed LPCGMN to train an epoch decreases from about 3 m 32 s to about 1 m 8 s when $L$ increases from 3 to 5. The results validate that the high-resolution geometric location map is decomposed by an LP with a reasonable number of layers and fed to the designed sub-network, respectively, which can sacrifice part of the prediction gain in exchange for expensive computational overhead.

\begin{figure}[!b]
	\centering  
	\includegraphics[scale=0.536]{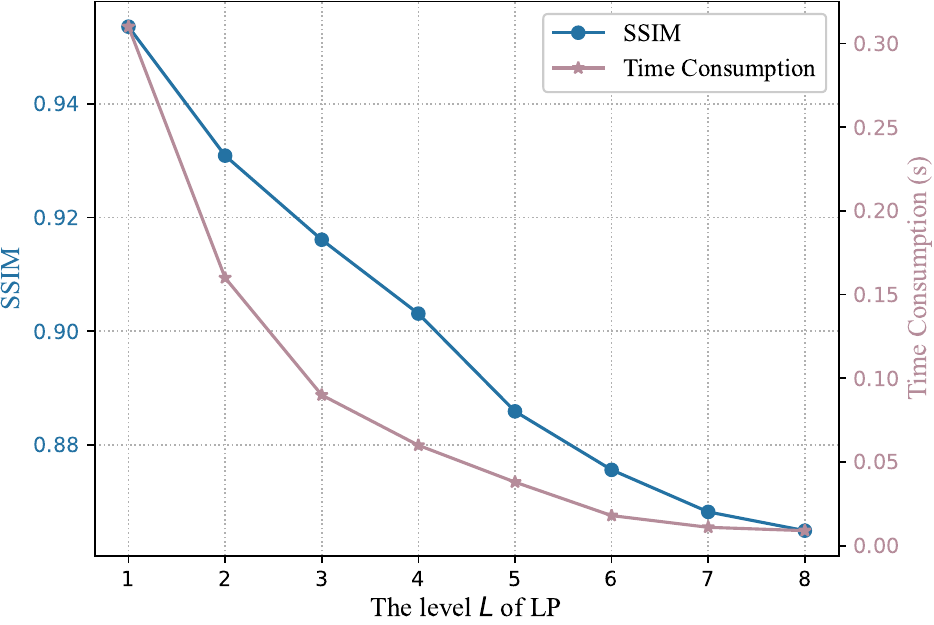}
	\captionsetup{font=footnotesize}
	\caption{Comparison of SSIM and time consumption (in seconds) of the proposed LPCGMN with different $L$ on DPM.}
	\label{fig:ssim}
\end{figure}

To comprehensively explore the influence of the proposed LPCGMN with different LP levels on CGM reconstruction performance, we compare the proposed LPCGMN with $L= 1,2,3,4,5,6,7,8$ under the two indicators of SSIM and inference time consumption. As shown in \figref{fig:ssim}, each result represents the average of 30 tests of the corresponding LPCGMN on DPM. Regarding SSIM performance, there is a decreasing trend as the value of $L$ increases. It is worth noting that there is also a significant decrease in terms of time consumption. Specifically, when $L$ increases from $1$ to $8$, the corresponding time consumption decreases from 0.31 s to 0.009 s. It highlights a non-trivial trade-off between time consumption and reconstruction accuracy for the proposed LPCGMN. Therefore, we believe that the proposed approach holds promising potential for enabling future ultra-high-resolution CGM reconstruction, even within the constraints of demanding computational costs and real-time requirements.
\begin{figure*}[!t]
	\centering
	\subfloat[LPCGMN, $L = 3$.]{\includegraphics[width=0.33\textwidth]{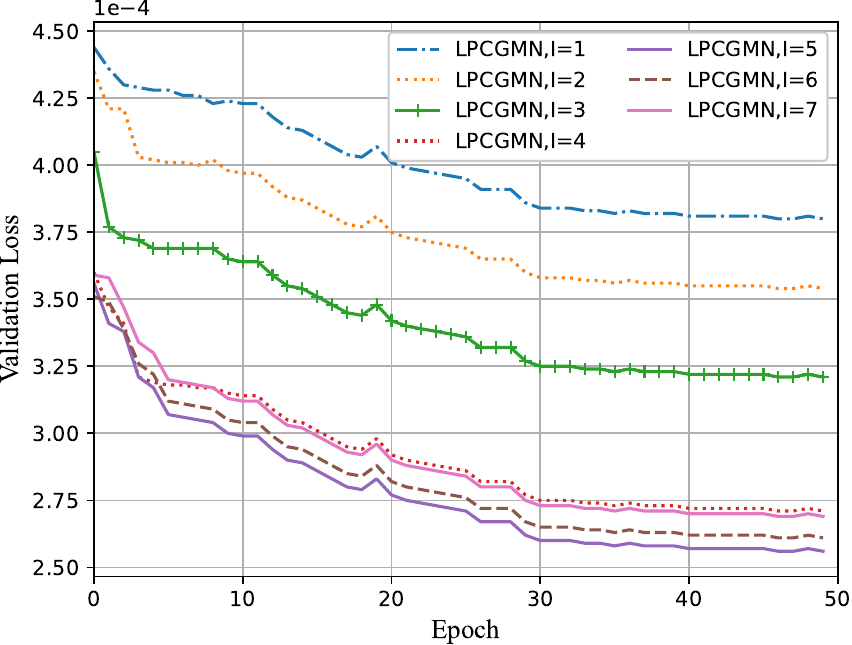}\label{fig:lossa}}
	%\hspace{0.5in}
	\hfill
	\subfloat[LPCGMN, $L = 4$.]{\includegraphics[width=0.33\textwidth]{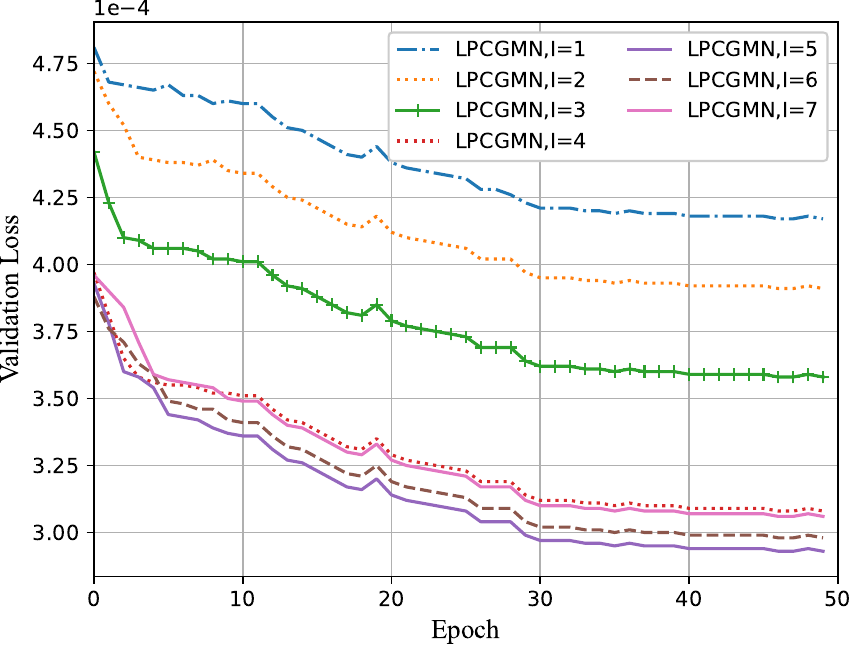}\label{fig:lossb}}
	%\hspace{0.5in}
	\hfill
	\subfloat[LPCGMN, $L = 5$.]{\includegraphics[width=0.33\textwidth]{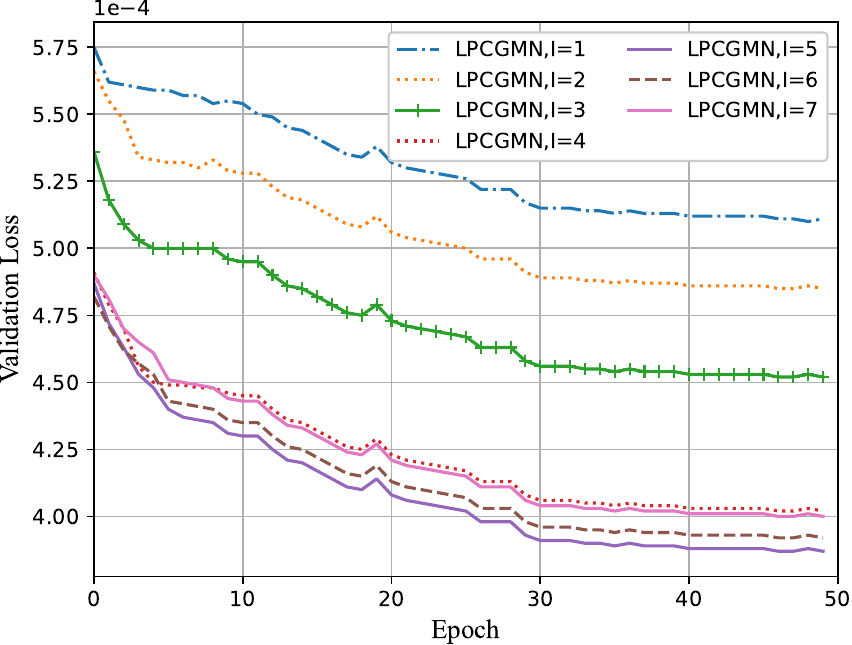}\label{fig:lossc}}
	\captionsetup{font=footnotesize}
	\caption{The impact of the number of attention heads on the validation loss of the LPCGMN on DPM under three parameter settings of $L=3$, $L=4$, and $L=5$.}
	\label{fig:loss}
\end{figure*}

\subsubsection{Experiment Results on LPCGMN with Different Numbers of Attention Heads}
Further, we need to determine the effect of the number $I$ of attention heads in the MHCCA and MHSA modules on the ability of the LPCGMN to predict channel gain. The MHCCA and MHSA use the attention mechanism to incorporate global environment information into the local features. However, increasing the number of attention heads in the attention mechanism results in a corresponding increase in model parameters. Although more attention heads can bring richer learning subspaces so the model can perform comprehensive scoring, the costly computation involved in multi-head attention modules impedes the development of lightweight and fast inference models. To this end, we evaluate the number of attention heads, $I = [1,2,3,4,5,6,7]$, in the two modules of the proposed LPCGMN under the setting of the number of LP levels, $L = [3,4,5]$, respectively. The accuracy of the reconstructed CGM serves as a critical performance metric. Therefore, at the end of each training epoch, the proposed LPCGMN saves the current optimal model weights, complete the model validation, and record the corresponding validation loss, namely MSE. As shown in \figref{fig:loss}, the influence of the number of attention heads on the performance of the LPCGMN to predict the channel gain is shown under the three parameter settings of the LP levels, $L = 3$, $L = 4$ and $L = 5$, respectively. It can be found that when the number of attention heads increases from 1 to 5, the proposed LPCGMN is robust when increasing the number $I$ of attention heads to reduce the prediction error. Specifically, as shown in \subfigref{fig:loss}{fig:lossa}, compared to the LPCGMN with $I=1$, the LPCGMN with $I=5$ increases the reconstruction gain by 1.72 dB. However, we are surprised to observe that when $I$ is set to be greater than 5, the prediction error of the proposed LPCGMN does not decrease but slightly increases. The LPCGMN with $I=6$ and $I=7$ reduces the performance gain of 0.66 dB and 0.22 dB, respectively, compared with the LPCGMN with $I=5$. We think that this may be due to the size of image tested is not large enough, but when the number of learning subspaces is too large, more subspaces lose their mutual independence, resulting in the overlapping phenomenon between them, which affects the effective learning of the LPCGMN, and the stagnation state appears. Besides, according to \figref{fig:loss}, it is still found that the LP module trades the prediction error for the overhead of time computation. Regarding the number of attention heads, $I=5$, the prediction performance of the LPCGMN decreases by about 1.80 dB from $L=3$ to $L=5$. Note that $I$ is set to 5 by default for the proposed LPCGMN in subsequent experiments if not specified.
\newcolumntype{L}{>{\hspace*{-\tabcolsep}}l}
\newcolumntype{R}{c<{\hspace*{-\tabcolsep}}}
\definecolor{lightblue}{rgb}{0.93,0.95,1.0}
\begin{table*}[!b]
	\captionsetup{font=footnotesize}
	\caption{The NMSE, RMSE and Storage Complexity Performance Comparison of the Proposed LPCGMN and the Baselines in Different Scenarios}\label{Performance comparison of the proposed LPCGMN and other schemes in different scenarios}
	\centering
	\setlength{\tabcolsep}{1.7mm}%宽度
	\ra{1.7}%高度
	\scriptsize
	\scalebox{0.8}{\begin{tabular}{LcccccccccccccccccR}
			\toprule
			\multirow{2}{*}{Method}        & \multicolumn{2}{c}{DPM}                                 & \multicolumn{2}{c}{DPMcars}                             & \multicolumn{2}{c}{IRT2}                                & \multicolumn{2}{c}{IRT2cars}                            & \multicolumn{2}{c}{\begin{tabular}[c]{@{}c@{}}Zero Shot\\ IRT4\end{tabular}} & \multicolumn{2}{c}{\begin{tabular}[c]{@{}c@{}}Zero Shot\\  IRT4cars\end{tabular}} & \multicolumn{2}{c}{IRT4}                               & \multicolumn{2}{c}{IRT4cars}                            & \multirow{2}{*}{\begin{tabular}[R]{@{}R@{}}Model Size\\  Params\end{tabular}}  \\
			& \multicolumn{1}{c}{NMSE}   & \multicolumn{1}{c}{RMSE}   & \multicolumn{1}{c}{NMSE}   & \multicolumn{1}{c}{RMSE}   & \multicolumn{1}{c}{NMSE}   & \multicolumn{1}{c}{RMSE}   & \multicolumn{1}{c}{NMSE}   & \multicolumn{1}{c}{RMSE}   & \multicolumn{1}{c}{NMSE}             & \multicolumn{1}{c}{RMSE}              & \multicolumn{1}{c}{NMSE}                & \multicolumn{1}{c}{RMSE}                & \multicolumn{1}{c}{NMSE}  & \multicolumn{1}{c}{RMSE}   & \multicolumn{1}{c}{NMSE}   & \multicolumn{1}{c}{RMSE}    \\
			\midrule
			\rowcolor{lightblue}
			Unet \cite{ronneberger2015u}                           & \multicolumn{1}{c}{0.0087} & \multicolumn{1}{c}{0.0214} & \multicolumn{1}{c}{0.0077} & \multicolumn{1}{c}{0.0193} & \multicolumn{1}{c}{0.0301} & \multicolumn{1}{c}{0.0430} & \multicolumn{1}{c}{0.0292} & \multicolumn{1}{c}{0.0425} & \multicolumn{1}{c}{0.029}            & \multicolumn{1}{c}{0.0387}            & \multicolumn{1}{c}{0.0190}              & \multicolumn{1}{c}{0.0299}              & \multicolumn{1}{c}{0.0162} & \multicolumn{1}{c}{0.0291} & \multicolumn{1}{c}{0.0181} & \multicolumn{1}{c}{0.0293} &\multicolumn{1}{R}{\textbf{13.27M}}                  \\
			Wnet \cite{9354041}                          & 0.0075                     & 0.0200                     & 0.0060                     & 0.0170                     & 0.0219                     & 0.0320                     & 0.0182                     & 0.0286                     & 0.0284                               & 0.0384                                & 0.0186                                  & 0.0296                                  & 0.0135                    & 0.0262                     & 0.0144                     & 0.0261                     &\multicolumn{1}{R}{26.54M}                                 \\
			\rowcolor{lightblue}
			\multicolumn{1}{L}{LPCGMN, $L=3$ $\downarrow$} & \textbf{0.0064}            & \textbf{0.0160}            & \textbf{0.0057}            & \textbf{0.0152}            & \textbf{0.0188}            & \textbf{0.0211}            & \textbf{0.0135}            & \textbf{0.0203}            & \textbf{0.0240}                      & \textbf{0.0312}                       & \textbf{0.0147}                         & \textbf{0.0205}                         & \textbf{0.0125}           & \textbf{0.0187}            & \textbf{0.0134}            & \textbf{0.0189}            & \multicolumn{1}{R}{14.8M}                                     \\
			\multicolumn{1}{L}{LPCGMN, $L=4$} & 0.0068                     & 0.0171                     & 0.0059                     & 0.0163                     & 0.0247                     & 0.0325                     & 0.0140                     & 0.0218                     & 0.0257                               & 0.0334                                & 0.0156                                  & 0.0269                                  & 0.0131                    & 0.0196                     & 0.0148                     & 0.0263                     &\multicolumn{1}{R}{17.2M}                                                               \\
			\rowcolor{lightblue}
			\multicolumn{1}{L}{LPCGMN, $L=5$} & 0.0078                     & 0.0204                     & 0.0070                     & 0.0182                     & 0.0222                     & 0.0323                     & 0.0159                     & 0.0271                     & 0.0268                               & 0.0348                                & 0.0166                                  & 0.0292                                  & 0.0147                    & 0.0267                     & 0.0153                     & 0.0269                     & \multicolumn{1}{R}{20.1M}                                                                        \\
			\bottomrule
		\end{tabular}
	}
\end{table*}                  
%进一步，我们需要确定MHCCA和MHSA模块中注意力头数I对LPCGMN预测信道路径增益能力的影响。MHCCA和MHSA模块takes advantage of the attention mechanism to encode longrange global information into the local features.然而，注意力机制仍然存在一个权衡关系在注意力头数和模型参数量之间，虽然更多的注意力头数可以带来更为丰富的学习子空间，使得模型可以进行全方位的打分，但是多头自注意力模块的昂贵计算阻碍了轻量级和快速推理模型的设计。为此，我们分别在拉普拉斯金字塔层数L=[3,4,5]的设置下，对LPCGMN的两个模块中注意力头数I=[2,3,4,5,6,7,8]进行了评估。The accuracy of the reconstructed CGM is a crucial performance evaluation indicator. 所以在LPCGMN训练的每一个epoch结束时，我们会保存当下最优的模型权重并完成一次模型的验证，记录下对应的验证损失，即MSE。如图8所示，分别展示了当L=3，L=4和L=5三个参数设置下注意力头数对LPCGMN模型预测能的影响。从中可以发现，当注意力头数从1增加到5的期间，the proposed LPCGMN is robust when increasing the attention heads I to reduce the predict error.具体而言，如图8（a）所示，I=5的LPCGMN相较于I=1的LPCGMN增加了1.72dB的性能增益。然而，我们惊讶的发现当I设置为大于5时，LPCGMN预测的误差并没有减小，反而稍有增加，其中I=7的LPCGMN相较于I=5的LPCGMN减少了0.22dB的性能增益。我们分析认为这是因为学习的子空间数目过大时，会出现多个空间丧失相互的独立性，生成学习子空间的重叠现象，从而影响LPCGMN的有效学习，出现了停滞的状态。此外，根据图8，仍然可以发现拉普拉斯金子塔模块是以预测误差来换取模型计算的开销。就注意力头数I=5而言，LPCGMN的预测性能从L=3到L=5，下降了大概1.80dB。
%\begin{figure*}[!t]
%\centering
%\subfloat[LPCGMN, $L = 3$.]{\includegraphics[width=0.33\textwidth]{figure/l=3MHnew.eps}\label{fig:lossa}}
%%\hspace{0.5in}
%\hfill
%\subfloat[LPCGMN, $L = 4$.]{\includegraphics[width=0.33\textwidth]{figure/l=4MHnew.eps}\label{fig:lossb}}
%%\hspace{0.5in}
%\hfill
%\subfloat[LPCGMN, $L = 5$.]{\includegraphics[width=0.33\textwidth]{figure/l=5MHnew.eps}\label{fig:lossc}}
%\captionsetup{font=footnotesize}
%\caption{The impact of the number of attention heads on the validation loss of the LPCGMN on DPM under three parameter settings of $L=3$, $L=4$, and $L=5$.}
%\label{fig:loss}
%\end{figure*}
\subsubsection{Experiment Results on Performance Comparisons}
One benchmark is the UNet, which has a U-shaped encoder-decoder structure, and its encoder and decoder are two symmetric CNNs. The encoder part is responsible for extracting features from the input image at a lower spatial resolution, and the decoder part recovers the feature map to the original resolution. The WNet (U + U makes a W) is another benchmark that predicts channel gain by combining two UNets. To ensure a fair comparison, we employed a uniform training strategy for all neural networks, and the experimental results are shown in Table II.
\begin{figure}[!t]
\centering
\includegraphics[scale=0.61]{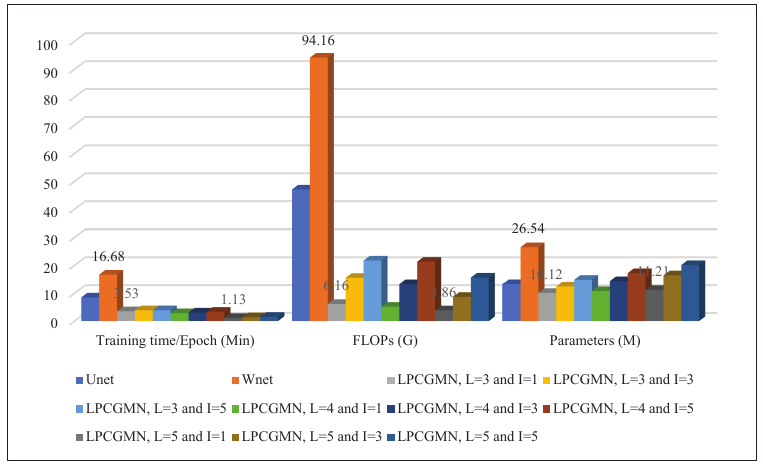}
\captionsetup{font=footnotesize}
\caption{Performance comparison of the proposed LPCGMN and the baseline methods on DPM.}
\label{fig:flops}
\end{figure}

\textit{i) The Accuracy of Reconstructed CGM.} In order to conduct a comprehensive comparison between our model and UNet and WNet, we train and test on several different scenarios: DPM, DPMcars, IRT2, IRT2cars, IRT4 and IRT4cars, respectively. The reconstruction accuracy is all given in the form of NMSE and RMSE. RMSE is calculated as the square root of the MSE over all test samples. Table II compares the reconstruction accuracy of different methods under the six types of communication scenario datasets. According to Table II, our proposed method achieves lower NMSE and RMS compared to the baselines. Note that our proposed LPCGMN can achieve optimal performance when $L$ equals 3 and $I$ equals 5. By increasing the parameter $L$, the proposed LPCGMN is capable of continuously reducing the overall computational burden, which is further analyzed in subsequent experiments.

\textit{ii) Generalization and Complexity Analysis.} The ability to generalize refers to whether the learned model can maintain a competitive prediction accuracy on unknown (unseen) datasets. Zero-shot learning is one of the key ways to evaluate the generalization performance. In our experiments, Zero-shot IRT4 and Zero-shot IRT4cars mean testing these methods, which are trained on DPM, on IRT4 or IRT4cars. Specifically, we train two benchmarks (Unet and Wnet) and three LPCGMN with different numbers of LP layers, respectively, on the DPM dataset. According to Table II, our proposed LPCGMN with $L=3$ has lower NMSE and RMSE on Zero-shot IRT4 and Zero-shot IRT4cars, which means that the proposed LPCGMN has competitive generalization ability and learning transfer ability.

The complexity of a neural network model is mainly determined by two key indicators: The number of model parameters and floating point operations (FLOPs). In addition, the training time per epoch is also important, which means the cost of model training and deployment. To be specific, we visualize the model parameters, FLOPs, and training time per epoch during training on the DPM dataset for two benchmarks and nine LPCGMN models with different numbers of layers and attention heads, as shown in \figref{fig:flops}. According to \figref{fig:flops}, in terms of training time per Epoch, Wnet requires a high training duration, which is about 16.68 minutes. The proposed LPCGMN with $L=5$ and $I=1$ requires the least training time, roughly 1.13 minutes. In terms of FLOPs, Wnet is still the highest, which is about 94.16 Giga FLOPs. The proposed LPCGMN with L=3 and I=1 and LPCGMN with L=5 and I=1 are 6.16 and 3.86 Giga FLOPs, respectively. According to \figref{fig:loss} and \figref{fig:flops}, we further verify that there is a trade-off between reconstruction accuracy and time consumption. Besides, the proposed LPCGMN is robust when increasing the level $L$ of LP to reduce FLOPs. Regarding model parameters, the lightest model is the LPCGMN with L=3 and I=1, which is about 10.12 million, while Wnet has a large number of model parameters, roughly 26.54 million.

\section{Conclusion}\label{sec:conc}
Our study investigated how to efficiently construct CGM, which is an essential enabler for environment-aware communication. To overcome the limitations of existing SM-based and DL-based channel knowledge estimation methods, we cast the CGM construction problem as a task of I2I reconstruction. In our work, rather than directly employing a neural network for constructing CGM, as done in the previous works, we introduced a parameter-free LP to replace traditional disentanglement and reconstruction frameworks, thereby significantly enhancing the inference speed. Specifically, we identified that the differences in low-frequency components are more pronounced in the dataset we utilized. According to this observation, we designed specialized subnetworks for low- and high-frequency components, respectively. More importantly, to make our model lightweight and enhance accuracy, we proposed three essential modules to reduce model parameters and strengthen the local-global feature interaction. Our results showed the effectiveness and generalization capability of the proposed method. Furthermore, the proposed approach demonstrated significant potential to tackle future ultra-high-resolution CKM reconstruction tasks, particularly in scenarios characterized by limited computing resources and demanding real-time requirements. Hence, our work shows promise in developing a lightweight, low-latency, and high-accuracy CKM reconstruction model for future wireless communication networks.

\bibliographystyle{IEEEtran}
\bibliography{EE_AI}

\end{document}